\documentstyle[12pt,aasms4]{article}
\lefthead{Schmidt et al.}
\righthead{The High-Z SN Search}
\begin{document}
\title{The High-Z Supernova Search: Measuring Cosmic Deceleration and Global Curvature of the Universe Using Type Ia Supernovae\footnote[1]{This work is based in part on observations at the European Southern Observatory, La Silla, Chile.}\\
({\it To appear in the Astrophysical Journal})}
\author{Brian P. Schmidt\altaffilmark{2}, Nicholas B. Suntzeff\altaffilmark{3}, M. M. Phillips\altaffilmark{3}, Robert A. Schommer\altaffilmark{3}, Alejandro Clocchiatti\altaffilmark{3,4}, Robert P. Kirshner\altaffilmark{5}, Peter  Garnavich\altaffilmark{5}, Peter Challis\altaffilmark{5}, B. Leibundgut\altaffilmark{6}, J. Spyromilio\altaffilmark{6}, Adam G. Riess\altaffilmark{5,7}, Alexei V. Filippenko\altaffilmark{7}, Mario Hamuy\altaffilmark{8}, R. Chris Smith\altaffilmark{4,9}, Craig Hogan\altaffilmark{10}, Christopher Stubbs\altaffilmark{10}, Alan Diercks\altaffilmark{10}, David Reiss\altaffilmark{10}, Ron Gilliland\altaffilmark{11}, John Tonry\altaffilmark{12}, Jos\'e Maza\altaffilmark{13}, A. Dressler\altaffilmark{14},  J. Walsh\altaffilmark{6}, and R. Ciardullo\altaffilmark{15}}
\altaffiltext{2}{Mount Stromlo and Siding Spring Observatories, Private Bag, Weston Creek P.O., ACT 2611, Australia.}
\altaffiltext{3}{Cerro Tololo Inter-American Observatory, Casilla 603, La Serena, Chile; National Optical Astronomy Observatories, operated by the Association of Universities for Research in Astronomy, Inc. (AURA), under cooperative agreement with the National Science Foundation.}
\altaffiltext{4}{Current Address: Departamento de Astronomia y Astrofisica, Pontificia Universidad Catolica de Chile, Casilla 104, Santiago 22, Chile}
\altaffiltext{5}{Harvard-Smithsonian Center for Astrophysics, 60 Garden St., Cambridge, MA 02138.}
\altaffiltext{6}{European Southern Observatory, Karl-Schwarzschild-Strasse 2, D-85748 Garching, Germany.}
\altaffiltext{7}{Department of Astronomy, University of California, Berkeley, CA 94720-3411.}
\altaffiltext{8}{Steward Observatory, University of Arizona, Tucson, AZ 85721.}
\altaffiltext{9}{University of Michigan, Department of Astronomy, 834 Dennison, Ann Arbor, MI 48109-1090.}
\altaffiltext{10}{Department of Astronomy, University of Washington, Seattle, WA 98195-1580.}
\altaffiltext{11}{Space Telescope Science Institute, 3700 San Martin Drive, Baltimore, MD 21218.}
\altaffiltext{12}{Institute for Astronomy, University of Hawaii, Manoa, Honolulu, HI 96822.}
\altaffiltext{13}{Deparmento Astronom\'ia, Universidad de Chile, Casilla 36-D, Santiago, Chile.}
\altaffiltext{14}{Carnegie Observatories, 813 Santa Barbara Street, Pasadena, CA 91101.}
\altaffiltext{15}{Pennsylvania State University, Department of Astronomy \& Astrophysics, 525 Davey Laboratory, University Park, PA 16802.}
\begin{abstract}
The High-Z Supernova Search is an international collaboration to discover
and monitor type Ia supernovae (SN~Ia) at $z > 0.2$ with the aim of measuring cosmic
deceleration and global curvature. Our collaboration has pursued a basic understanding
of supernovae in the nearby Universe, discovering and observing a large sample of objects, 
and developing methods to measure accurate distances with SN~Ia. 
This paper describes the extension of this program to $z \geq 0.2$, outlining
our search techniques and follow-up program. We have devised high-throughput filters
which provide accurate two-color restframe $B$ and $V$ light curves of SN~Ia, enabling us
to produce precise, extinction-corrected luminosity distances in the range $0.25 < z < 0.55$.
Sources of systematic error from K-corrections, extinction, selection effects, and evolution are  
investigated, and their effects estimated.  We present photometric and spectral observations of
SN~1995K, our program's first supernova, and use the data to
obtain a precise measurement of the luminosity distance to the $z=0.479$ host galaxy. 
This object, when combined with a nearby sample of SN, yields an estimate for the matter density 
of the Universe of $\Omega_M = -0.2^{+1.0}_{-0.8}$ if $\Omega_\Lambda = 0$.
For a spatially flat universe composed of normal matter and a cosmological constant, we 
find $\Omega_M = 0.4^{+0.5}_{-0.4}$, $\Omega_\Lambda = 0.6^{+0.4}_{-0.5}$. We demonstrate that
with a sample of $\sim 30$ objects, we should be able to determine relative luminosity distances
over the range $0 < z< 0.5$ with sufficient precision to measure $\Omega_M$ with an uncertainty of $\pm 0.2$. 

\end{abstract}
\keywords{cosmology: observations --- galaxies: distances and redshifts --- supernovae: general --- supernovae: individual (SN 1995K)}
\section{Introduction}
Measuring the cosmological parameters which describe the global properties of the
Universe has been a fundamental quest in astronomy ever since Robertson \markcite{R36}(1936) and
Walker \markcite{W36}(1936) formulated the metric for a homogeneous and isotropic Universe.  
By observing how a standard candle dims as a function of redshift, usually shown as a Hubble
diagram, the effects of curvature and cosmic deceleration can be observed and quantified (\markcite{S61}Sandage 1961).
Early luminosity distance investigations (\markcite{HMS56}Humason, Mayall, \& Sandage 1956; \markcite{B57}Baum 1957; Minkowski 1960\markcite{M60}) used brightest cluster galaxies as standard candles and measured galaxy brightnesses in the range
$0.01 < z < 0.5$. Attempts to trace luminosity distances versus redshifts with these galaxies at $z>0.1$
changed emphasis when it was realized both from theory (\markcite{T72}Tinsley 1972) and observation (e.g., \markcite{OGH96}Oke, Gunn, \& Hoessel 1996) that the effects of galaxy evolution are much larger than the
differences due to cosmology. 

Although many other methods for measuring the global curvature and cosmic deceleration
exist (e.g., Peebles \markcite{Peebles93} 1993), none
has yet delivered a definitive result.  For example,
measuring the number
of galaxies as a function of magnitude maps out the volume of
space as a function of redshift, and can be used to gauge 
the global geometry. Attempts to use this method \markcite{S84}(Shanks et al. 1984) have been
hampered by galaxy evolution and merging, although some of the uncertainty may be
eliminated by moving to the infrared
(\markcite{yp95}Yoshi \& Peterson 1995). Another test examines the angular size of a standard
rod as a function of redshift. Kellerman \markcite{K93}(1993) resolved
a large sample of compact radio sources out to $z\approx3$ using Very
Long Baseline Interferometry. Although the angular sizes increase at $z> 1$
as expected for a non-empty Universe, evolutionary effects are hard
to quantify. Stepanas \& Saha \markcite{SS95}(1995) have also shown that 
the unknown intrinsic distribution of source sizes makes it difficult to obtain a
statistically significant measurement of cosmological parameters. Guerra \& Daly (1998)\markcite{GD98}
have also used extended radio galaxies as standard rods and show 
that the results are consistent with a low density universe.

Type Ia supernovae (SN~Ia) have long been considered promising tools
for measuring extragalactic luminosity distances, but only recent searches, the
resulting sets of light curves and spectra, and new methods of analysis (\markcite {P93}Phillips 1993 [P93]; \markcite{H96a}Hamuy et al.
1995, 1996a-d [H95,H96a-d]; \markcite{RPK95} Riess, Press, \& 
Kirshner 1995, 1996a [RPK95, RPK96]) have quantified the
nature, power, and limitations of SN~Ia as distance indicators.
SN~Ia offer high intrinsic luminosity [$M_B\approx-19.4$~mag (\markcite{Setal97}Saha et al. 1997)] and
as individual stars, may not be subject to the same evolutionary effects which plague galaxies (although
this must be demonstrated). Even
before this breakthrough in
understanding SN~Ia, two searches for distant
SN~Ia were initiated (\markcite{NN89}N\o rgaard-Nielsen et al. 1989; \markcite{Pe95}
Perlmutter et al. 1995) to measure cosmological parameters. These searches
demonstrate that it is possible to find SN at $z > 0.3$ with large-format CCD detectors, and give hope
that a significant sample of SN~Ia can be gathered in just a few years.
Perlmutter et al. (1997, \markcite{Pe98}1998) have already presented observations of 8 objects at $z > 0.35$.
A  sample of $\sim 30$ objects, if carefully measured and shrewdly analyzed,  will provide a statistically
interesting measurement of  global cosmological parameters. 

Recently, much effort has been put into examining how to use the power spectrum of 
fluctuations in the cosmic microwave background (CMB) to measure cosmological parameters
(see\markcite{Hu96} Hu 1996 for a review). Future satellite missions, such as the Microwave Anisotropy Probe and
Planck Missions,
will measure  temperature fluctuations across the sky on scales substantially less than a degree,
determining the power spectrum of fluctuations out to a multipole expansion of $\ell >500$. 
Model fits to these observations promise to provide simultaneous measurement of ten cosmological parameters. However,
because the fits are degenerate for certain combinations of interesting parameters such as $\Omega_M$ and $\Omega_\Lambda$, the CMB
observations will need to be combined with other observational data, such as those from high-redshift supernovae (\markcite{ZSS97}Zaldarriaga, Spergel, \& Seljak 1997), to determine $\Omega_M$ and $\Omega_\Lambda$ individually. 

This paper reports on the High-Z SN Search, a coordinated program to discover,
spectroscopically classify, and measure photometrically in at least two filters
a distant set of SN~Ia. This enterprise aims to measure the deceleration parameter, $q_0$, 
with an uncertainty smaller than $0.1$, and will place strong limits on combinations of cosmological
parameters such as $\Omega_M - \Omega_\Lambda$.
The CMB observations provide a nearly orthogonal set of parameters, so we will be 
able to separate the effects of any exotic forms of matter-energy in the Universe from
normal matter.  We will be able to discern whether the Universe is open, closed, or has 
nearly zero global curvature. Preliminary results from our program were reported by 
Schmidt et al. (\markcite{S96}1996), Schmidt \markcite{S97}(1997), and Leibundgut \& Spyromilio
(1997)\markcite{LS97}. In addition, we have confirmed the predicted time-dilation of redshifted
objects using SN~Ia light curves (Leibundgut et al. 1996\markcite{L96}) and spectra
(Riess et al. 1997 \markcite{retal97}), and presented observations of 3 objects observed
with {\it HST} (Garnavich et al. 1998).

In \S 2 we describe how the expansion, deceleration, 
and curvature of the Universe are related to luminosity distances, and in \S 3 we discuss
measuring distances with SN~Ia at $z<0.1$. Our supernova search and observational
follow-up program are outlined in \S 4. In \S 5 we present the techniques and limitations
of using SN~Ia to measure accurate luminosity distances at $z>0.2$. Observations of the first
SN discovered by this program, SN~1995K at z=0.479, are presented  in \S 6, with the 
techniques discussed in \S 3-5 applied to this object. We summarize the High-Z SN Search to date in \S 7 and
use the results for SN~1995K to estimate the precision with which we will be able to constrain
cosmological parameters.

\section{Expansion, Deceleration, and Curvature}

The precise large-scale isotropy of the microwave background confirms a picture
in which the Universe is accurately described on large scales by the maximally
symmetric, Robertson-Walker line element (e.g., \markcite{W72} Weinberg 1972).
For events with time separation $dt$, radial coordinate separation
$dr$, and angular separation $d\theta$, the line element $ds$ is given by
\begin{equation}
ds^2=dt^2-a^2(t)\left[{dr^2\over{1-kr^2}}+r^2d\theta^2\right].\label{eq:RW}
\end{equation}
The global spatial geometry has
the character of a hypersphere of radius $k^{-1/2} a(t)$, where
$a(t)$ is the cosmic scale factor which defines 
the physical scale of the hypersphere at each time.
In these units  the spatial curvature parameter
$k$ can be 1, 0, or $-1$, corresponding to a closed, flat, or open Universe, respectively.

The complete spacetime metric, which depends on $a(t)$, is 
determined by the Friedmann equation
\begin{equation}
H^2\equiv(\dot a/a)^2={8\pi G\rho\over 3} - {k\over a^2}, \label{eq:F}
\end{equation}
where $\rho$ is the total density of all forms
of matter-energy. Friedmann-Robertson-Walker (FRW) cosmologies are based on  equations (\ref{eq:RW}) and (\ref{eq:F}),
and provide a complete description of an isotropic and homogeneous universe.

We adopt a conventional model in which  the matter content  of
the Universe is composed of a sum of components each having a fraction $\Omega_i$ of the 
current critical density $\rho_{crit}\equiv 3H_0^2/8\pi G$ and various equations  of state with density $\rho_i\propto
({\rm volume})^{-(1+\alpha_i)}$ [e.g., $\alpha=0$ for normal matter ($\Omega_M$), $\alpha=-1$ for a
cosmological constant ($\Omega_\Lambda$), $\alpha=+1/3$ for radiation ($\Omega_{\rm rad}$),
$\alpha=-1/3$ for non-commuting strings ($\Omega_{\rm S}$)].
It is  convenient to adopt a   parameter $\kappa_0\equiv 
k c^2/[a(t_0)^2H_0^2]$ representing  the  scalar curvature   in units commensurate with
the density parameters;  the current physical  radius of hypersphere curvature is 
$k^{-1/2} a(t_0)= \kappa_0^{-1/2} c  H_0^{-1}$ and the definition of critical density gives
$\kappa_0=\sum_i\Omega_i-1$. We can then write the Friedmann equation in terms 
of these model parameters, 
\begin{equation}
H^2= H_0^2[ \sum_i\Omega_i(1+z )^{3+3\alpha_i}-\kappa_0(1+z )^2].\label{eq:F1}
\end{equation}
It is conventional to define a  ``deceleration parameter'' $q_0\equiv -\ddot a(t_0) a(t_0)/\dot a^2(t_0)$,
characterizing the low-redshift behavior, which can be expressed as
\begin{equation}
 q_0= {1\over 2}\sum_i\Omega_i(3+3\alpha_i)-\sum_i\Omega_i ={1\over 2}\sum_i\Omega_i(1+3\alpha_i).\label{eq:q}
\end{equation}

Distance measurements based on SN~Ia light curves are described as
luminosity distances, $D_L$, and are defined  by the ratio of
the intrinsic luminosity $\cal L$ to the observed flux $\cal F$ as
\begin{equation}
D_{L} = \left(\frac{{\cal L}}{4 \pi {\cal F}}\right)^{\frac{1}{2}}.\label{eq:DL}
\end{equation}
In FRW cosmologies  $D_L$  is derived by computing the area of the
sphere over which the flux is distributed from a source at a radial
coordinate fixed by the redshift. Including the effects of time dilation 
and redshift, the luminosity distance is
\begin{equation}
 D_L H_0= (1+z)\left | \kappa_0 \right |^{-1/2}S\lbrace\left | \kappa_0 \right |^{1/2}
\int_0^zdz'[\sum_i
\Omega_i(1+z')^{3+3\alpha_i}-\kappa_0(1+z')^2 ]^{-1/2}\rbrace, \label{eq:intDL}
\end{equation}
where $S\{x\}\equiv \sin(x)$, $x$, or $\sinh(x)$  for
$k= 1,0,-1$, respectively (Coles \& Lucchin 1995\markcite{CL95}).

Mattig \markcite{M58}(1958) showed that when normal matter is the only contributor
to $\Omega$, 
\begin{equation}D_L H_0 = {1 \over q_0^2}\left[q_0z +(q_0-1)(\sqrt{1+2q_0z}-1)\right].\label{eq:mattig}
\end{equation}
Alternatively, equation (\ref{eq:intDL}) can be expanded in $z$  to
give
\begin{equation}
 D_L H_0= z+z^2\left({1-q_0\over 2}\right)  +{\cal O}(z^{3}), \label{eq:expansion}
\end{equation}
which is valid for all models. 
The linear term of the expansion is the Hubble law, and has been studied for
many years. Its linear form has been verified to high precision in the nearby
Universe using SN~Ia with the same techniques we employ for this
project (H96b; RPK96; Tammann \& Leibundgut 1990), using brightest cluster galaxies (Lauer \& Postman \markcite {LP92}1992), and
again using SN~Ia at larger redshifts (Kim et al. \markcite{K97}1997).
The current debate on the value of $H_0$ centers on obtaining an absolute calibration for these
distance indicators in nearby galaxies through accurate absolute distances.
The measurements of curvature and deceleration require only a relative distance indicator to
obtain the shape of the ($D_L$,$z$) relation, and are not affected by current uncertainties in $H_0$ due to
local calibration. Equation (\ref{eq:expansion}) shows that departures in the luminosity distance from a pure Hubble law, to lowest order
in $z$, are proportional to  $q_0$ ---  they depend only on deceleration and not on curvature.
With a distance modulus, $m-M = 5 \log_{10} \left(D_L/{\rm 10~parsec}\right)$, measured to precision $\Delta m$~mag for an object at redshift $z$, equation
(\ref{eq:expansion}) shows (using $\Delta m=5\Delta\log_{10}[H_0D_L]$) that we measure $q_0$ to a precision $\Delta q_0\approx 0.9 \Delta m/z$;
 thus a single well-observed SN~Ia at $z=0.5$ with $\sigma = 0.15$~mag (H96b, RPK96) should yield a
precision of about
$\Delta q_0=0.27$, almost a $2\sigma$ discrimination between an empty ($q_0=0$) and a flat
($q_0=0.5$) Universe. However, we caution the reader that already at $z=0.5$, ${\cal O}(z^{3})$ terms cannot be neglected, especially in cosmologies
with significant cosmological constants.

To illustrate the precise effects of cosmology on luminosity distance, we plot the 
differences in distance modulus, $m-M$, from an $\Omega_{\rm tot}=0$ Universe as a function of redshift 
for a set of universes composed of different amounts
and types of matter-energy (Figure \ref{fig:dlvsz}). Although first order deviations 
constrain only the linear combination of parameters corresponding to
deceleration, data on objects over a range of redshifts up to $z\approx 1$
can separate out the effects of the various forms of mass-energy
in the $(D_L,z)$ relation, and place limits on global curvature. In particular,
it is possible to separate flat cosmological models with non-zero
$\Omega_\Lambda$ from open universes containing only normal matter (Goobar and Perlmutter 1995).

\section{Using Type Ia Supernovae to Measure Luminosity Distances}	

SN~Ia have been used as extragalactic distance indicators since Kowal \markcite{K68}(1968)
published a Hubble diagram ($\sigma = 0.6$ mag) for SN~I.
We now recognize that the old SN~I spectroscopic class conflated two distinct
types of objects: SN Ib/c which are probably massive stars that undergo
core collapse after losing their hydrogen atmospheres, and SN~Ia which are most likely thermonuclear explosions
of white dwarfs (see Filippenko 1997 for a review). Modern versions of the SN~I (now SN~Ia) Hubble diagram shows
scatter about the inverse-square line of about $0.3$ to $0.5$ mag (\markcite{TL90}Tammann
\& Leibundgut 1990; \markcite{vBP92} van den Bergh \& Pazder 1992;
Branch \& Miller 1993\markcite{BM93}; Sandage \& Tammann 1993\markcite{ST93}), which is
remarkable given the heterogeneous sources and oftentimes poor observations upon which these diagrams
are based.		

The advent of precise observations of nearby SN~Ia made with CCD detectors produced evidence for
genuine differences in the luminosities, light curve shapes, and spectra among the Type Ia family.
SN~1984A (\markcite{B87}Branch 1987; \markcite{Betal89}Barbon, Rosino, \& Iijima 1989), SN~1986G (\markcite{P87}Phillips et al. 1987), SN~1991bg (\markcite{Fetal92}Filippenko et al. 1992b;\markcite{Letal93} Leibundgut et al. 1993), and SN~1991T
(\markcite{Fetal92}Filippenko et al. 1992a; \markcite{Petal92}Phillips et al. 1992) provided proof that SN~Ia were not all identical objects whose observed differences could be attributed to measurement errors, but that real differences among
these explosions are undoubtedly present.

The problem of understanding SN~Ia well enough to use them as cosmological probes despite their
intrinsic variation was solved by assembling a sufficiently large, uniform, and well-observed data set.
In 1990 a group of astronomers at CTIO and the University of Chile at Cerro Cal\'an 
initiated a systematic photographic search for SN~Ia using the Curtis Schmidt
telescope at CTIO (Hamuy et al. \markcite{H93} 1993a). Their program, which discovered 30 SN~Ia in $2.5$ years, also
acquired high-quality spectral and photometric follow-up for these supernovae.
The resulting data set (H96c) allows the precise determination
of the properties of SN~Ia as distance indicators. At maximum light, SN~Ia have an intrinsic range of
$>2$ mag in $B$ and $>1$~mag in $V$. Although this is an interesting
result for supernova physics, it does not bode well for using SN~Ia as high-precision distance
indicators without additional information.

Although their brightness at maximum light has a moderately large scatter, SN~Ia do exhibit a
correlation ($\sigma \approx 0.15$ mag) between
the rate at which their luminosity declines and absolute magnitude. P93 demonstrated this
relationship by plotting the absolute magnitude of ten nearby SN~Ia which
had dense photoelectric or  CCD coverage, versus the parameter $\Delta m_{15} (B)$, the amount by which
the SN decreased  in brightness in the $B$ band  over the 15 days following maximum light.
The sample showed a correlation, which when taken into account, dramatically
improved the predictive power of SN~Ia. The Cal\'an/Tololo survey yielded an independent confirmation of the P93
absolute magnitude-decline rate relationship from the sample of 30 SN Ia by using a $\chi^2$ fitting technique
to the $B,V$, and $I$ light curves (H95, H96a). When corrected for their rate of decline, H96c demonstrated that
the scatter in the Hubble diagram could be lowered to $\sigma \sim 0.15$~mag in $V$
for a sample of nearly 30 SN~Ia.  Another technique, the Multicolor Light Curve Shape (MLCS) method, has been
developed by  RPK95 and RPK96. By ``training'' on a nearby set of objects (P93's sample plus
a few additions), they achieve $\sigma < 0.2$~mag on a sample of 20 objects 
(H95, augmented by 10 additional well observed SN~Ia) in the Hubble flow.
This result is encouraging because the Hubble diagram derived by RPK96 
is independent of the objects on which their method was ``trained,''
and therefore provides an upper limit for the true dispersion of this distance measuring technique.
Other methods to correct for intrinsic luminosity differences or limits on the input sample by various
criteria have also been proposed to increase the precision of SN~Ia as distance indicators 
(\markcite{TS95}Tammann \& Sandage 1995;\markcite{F95} Fisher et al. 1995; \markcite{v95}van den Bergh 1995; 
\markcite{B96}Branch et al. 1996; \markcite{Pe97}Perlmutter et al. 1997).

The analyses described above assume that all SNe~Ia can accurately
be described by a one-parameter family of light curves. We know this is
not true because the scatter about the Hubble line in either H96c and RPK96 
is larger than the observational errors would indicate (H96c,RPK96). The inferred
scatter beyond the observational uncertainties is small, ($\sigma \approx 0.12$~mag), and the 
residuals (including observational uncertainties) are distributed about the mean with a
distribution consistent with a Gaussian. To this date, no other
observable has been  shown to successfully account for the remaining small intrinsic
scatter about the one parameter family of light curves. Unless supernovae are much
different at high redshift, the imperfection of SN~Ia as distance
indicators will have a negligible impact on using SN~Ia as cosmological probes.

\section{Search and Follow-Up Program}

Many techniques have been successfully used to discover supernovae, including
visual observations of nearby galaxies (\markcite{E94}Evans 1994), photographic surveys
(\markcite{Z68}Zwicky 1968; \markcite{M89}Mueller 1989, \markcite{Mc90}McNaught 1990, \markcite{H93a}Hamuy et al. 1993a; \markcite{P92}Pollas 1992), and CCD surveys
(\markcite{Pe92}\markcite{Pe95}Perlmutter et al. 1992, 1995; \markcite{T93} Treffers et al. 1993; \markcite{MWW97}Martin, Williams, \& Woodings 1997; \markcite{Reiss98}Reiss et al. 1998).
Although it is possible to discover objects up to $z\approx1$ (\markcite{S97c}Schmidt et al. 1997c)
by using large format CCDs coupled with wide fields on telescopes with the best image quality, it is 
efficient to measure cosmological parameters by observing objects in the range  $0.35 < z < 0.55$.
When systematic effects are small, the leverage gained with high-redshift objects is offset by
the difficulty in obtaining accurate measurements.
It is challenging to obtain accurate restframe $B$ and $V$ photometry of 
objects observed at $ 0.55 < z < 0.9$ because they are outside the optimum K-correction window
(Figure \ref{fig:Kcorrsig}), and these SN are currently less  powerful tools for measuring
cosmological parameters than their lower redshift siblings. The {\it Hubble Space Telescope}
({\it HST}) + WFPC2/NICMOS could acquire accurate restframe
$B$ and $V$ measurements for SN~Ia at $z\approx 1$. These objects hold the promise of establishing
powerful constraints on cosmology within this more distant observational window. From the
ground, however, the band $0.35 < z < 0.55$ gives the best combination of measurements and systematics
to investigate cosmology.

	\subsection{Observing Strategy}

To maximize the number of SN~Ia discovered in our target redshift range, $0.35 < z < 0.55$,
we observe a large area and aim to achieve a limiting magnitude of $m_R\approx23$~mag, 
which is $\sim 1$ magnitude fainter than the expected brightness of a $z=0.5$ SN~Ia at
maximum light (Figure \ref{fig:magvsz}). Finding objects is not the only
consideration; the objects must be found near or before maximum light, and we need to follow discoveries with spectra
and multi-color photometry. To ensure our objects are discovered young, we use the technique described
by Hamuy et al. (1993a) and Perlmutter et al. \markcite{Pe95}(1995), imaging fields near the end of a
dark run, and then reimaging the fields at the beginning of the next dark run. These two runs, separated by approximately 21 days
(close to the rise time of a time-dilated SN~Ia at z=0.5), provide objects which are at,
or before maximum light\footnote[16]{To target objects in the other advantageous K-correction window, $z\approx 1$,
observations to $m_I=24$~mag, separated by 30 days, will efficiently deliver young SN~Ia.}.
Observations are generally made near the celestial equator, so that we can use telescopes in both
hemispheres for follow-up spectroscopy and photometry. At least two observations are made at
each search position, to detect motion of asteroids, eliminate cosmic rays, and remove chip defects.

As an outgrowth of this project, Riess et al. \markcite{retal97}(1997) developed 
a method to measure the age of a SN~Ia relative to maximum light from its spectrum alone.
This technique is especially valuable because it provides another way to identify young objects, ideally
while still observing at the telescope.

Since we need to schedule large blocks of telescope time  months in advance to follow the supernovae, 
it is essential to have candidates after each supernova discovery run, and not be derailed
by weather. During the summer months of December through March, the Chilean Atacama desert has nearly 
100\% clear weather. We have concentrated our search efforts
at the CTIO 4~m telescope, the instrument which currently provides the widest field of view
of any large telescope in Chile.

On the CTIO 4~m telescope we image approximately $3$ square degrees per night
with a single 2048$^2$ detector, taking two consecutive 150~s $B45$ exposures of each field.
In good conditions ($1^{\prime\prime}$ seeing),
a combined frame has a limiting magnitude of $m_R\approx 23$, and provides a sufficiently long time baseline
to remove Kuiper belt objects, which have a typical parallax motion of $3^{\prime\prime}~{\rm hr}^{-1}$. Since
January 1997, the ``Big Throughput Camera'' (BTC) has been available at the CTIO~4~m telescope. This mosaic of
4 chips quadruples the imaging area, but has a somewhat longer readout time. We have recently used the BTC
to obtain two consecutive 300s $R$ exposures at every pointing, enabling us to cover 7 square degrees per night to a depth of $m_R=23.5$~mag.
	
	\subsection{Search Software}

Our supernova search is automated, with final cuts on potential candidates being made by eye. The
automated processing program is written in PERL, and calls IRAF tasks, DoPHOT
\markcite{SMS93}(Schechter, Mateo, \& Saha 1993), and various programs written in C. In brief, the program
aligns the second-epoch image with the first, initially finding the bright stars using DoPHOT, and
then matching  stars in the two frames using a triangle-matching algorithm similar to
that described by \markcite{G86}Groth (1986). The images are then 
registered using the IRAF tasks ``geomap'' and ``geotran.'' After registration, we match the point spread functions (PSF) of the two
epochs applying the method of Phillips \& Davis \markcite{PD95}(1995), which computes a convolution kernel
in Fourier space, and fits the high frequencies with a Gaussian profile. The DoPHOT analytic PSF
measurements show which image needs to be degraded, and indicates if the PSF matching 
cannot be made in a single convolution. This is the case when the images are elongated with respect to
each other such that neither image has a FWHM which is smaller than the other at all position angles.
In these cases, we convolve one of the images with an appropriate Gaussian, and then apply
the Phillips \& Davis method. After PSF matching, the images are scaled to the same intensity and sky
brightness values by plotting the intensity of each pixel in one image against the intensity of the
corresponding pixel in the other in a subraster centered on a star. We fit for an offset
(difference in sky brightness) and a scale (differential atmospheric transparency), and then
subtract the  intensity-scaled images. This procedure is carried out on both second-epoch images,
and these differenced images are averaged, rejecting any high pixels which are discordant by more than 3$\sigma$.

The resulting image is  searched using a point-source detection algorithm. Our algorithm samples the
combined difference image at many locations over the image, estimating the average noise
within a PSF, and then scans the image for objects above this threshold by a certain number of
$\sigma$ ($\sigma > 4$ being a typical choice).  A list of candidates, eliminating those near known
bright stars, is sorted by magnitude. This entire process takes about
6 minutes to run on a 170 MHz Sun UltraSparc for a pair of $2048^2$ images.  For inspection of candidates, the
examiner is presented with subrasters of the candidate's region from the first epoch, both
second epochs, and the subtracted image. These images can be viewed
simultaneously as a mosaic, or stacked and blinked.

The number of candidates to examine depends on the detection threshold and the quality of the match
between the two epochs. Typically, 5-50 objects are examined on each pair of search images, but most
are easily eliminated by inspection. Our approach is to minimize false alarms from a night of observing.
We usually have 5--20 possible SN candidates per night that are detected by our software
filter and which are not discarded by visual inspection. When there is doubt about the reality of 
a candidate, we make a repeat observation of the field. At this point, the candidate list is sent to
collaborators for
spectroscopic observations that can show whether or not the object is a supernova, give some information
on its type and age, and provide the redshift. Roughly 75\% of these candidates are confirmed as
SN~Ia, the remainder consisting of other supernova types, AGNs, and occasional mystery objects.
These mystery objects typically have no visible host galaxy, and fade by more than two magnitudes
within 24 hours (and in one case, at least 2 magnitudes in 3 hours). It is conceivable that these are
flare stars in the halo of the Milky Way, or the unbeamed optical counterparts to gamma-ray bursts
(Rhoads 1997\markcite{Rh97}). There is no bias in our selection against SN in which there is no
visible galaxy since the whole CCD field is searched.

	\subsection{Data Reduction Procedures}

We extract the high-redshift photometry in the same way we have measured the  $z< 0.1$ sample (H96c, RPK96).
We first calibrate a local photometric sequence of stars which appear in the CCD field of each supernova.
These stars are calibrated by observing standard stars on photometric nights, deriving color and
atmospheric extinction transformations, and then applying these to the local sequence.
The local photometric sequence, which typically spans a
substantial range of color and brightness, is then combined, correcting
for any color term of the system used in the observation, with relative photometry between the supernova
and sequence stars to produce a standard magnitude for the supernova. In general, the color terms of our
different systems are small, since most of the supernova observations are taken with identical filters as
described in Appendix A.

To produce precise relative photometry for our high-redshift supernovae we follow the same procedure
employed in the Cal\'an/Tololo survey for galaxy subtraction (Hamuy et al. 1994\markcite{H94}).
A template image, in which the supernova is absent, is required for every object/filter combination.
Ideally, these images would have better seeing than any of the other observations, so that the
observations are not degraded in the PSF matching process, and  should have more than twice the
signal-to-noise (S/N) ratio to minimize the addition of shot noise in the subtractions. In most
cases images of our SN discovery regions which are appropriate for use as templates are not in hand
before the time of
explosion, so we must return to these fields after about a year to obtain templates. One of our most difficult tasks is to obtain
good seeing images with long integration times to serve as acceptable templates. The analysis of our edata
set will be continuously improved as we build up improved templates for the supernovae we
follow.

We have developed a pipeline closely related to our search software for extracting photometry from our CCD frames. After normal processing in which the frames are bias subtracted and flat fielded, this pipeline
uses the search software to align an image to the template, match the two images' PSFs, scale the
intensity of the observation to the template, and then subtract the template from the observation
in a region around the galaxy. The software then runs DoPHOT on the galaxy-subtracted frame, using a fixed
position of the supernova\footnote[17]{In low S/N ratio situations, allowing the centroid of the PSF
to be a free parameter would cause a systematic over-estimate of the object's flux.} and a sequence
of comparison stars, to extract the magnitude of the supernova.  
After measuring the SN's brightness, the script builds a PSF from the comparison stars using DAOPHOT
(\markcite{Stet87}Stetson 1987), and
adds this PSF to user specified galaxies in the image so that they have the same brightness as the SN
measurement. After subtracting the template from the areas containing the added PSFs, DoPHOT is once
again run to extract the magnitudes of these artificial SN. These synthetic objects provide an effective
method to estimate the relative photometric errors, and can give a strong indication if something has
gone wrong in the reduction process.

Our spectra are reduced in a typical manner, except that we generally extract only a  small
region along the slit at the position of the supernova to minimize host galaxy contamination. We also
bin the spectra as necessary in the dispersion direction, after final reduction, to increase the S/N ratio per pixel, to aid in classifying the objects.

\section{Measuring Accurate Relative Distances Between Nearby and High-Z SN~Ia}	

The distances derived to SN~Ia are well characterized and tested in the nearby Universe,
but accurately comparing these objects to their more distant counterparts requires
great care. Selection effects can introduce systematic errors as a function
of redshift, as can uncertain K-corrections, or an evolution of the SN~Ia progenitor
population as a function of look-back time. These effects, if they are large, will
limit our ability to measure the ($D_L$,$z$) relation accurately, and have the potential
to sap the potency of high-redshift SN~Ia for measuring cosmological parameters. 

\subsection{Extinction}

Most attempts at using SN~Ia as distance indicators have not corrected
for extinction in the supernova host galaxy. The majority of the SN~Ia discovered
at $z<0.1$ which comprise the current sample of well observed
SN~Ia have been detected using photography at Schmidt telescopes, and it is
possible that many SN~Ia embedded in dusty spiral arms  or galaxy nuclei were missed by
these searches. Distant SN~Ia are discovered with CCD
observations  by subtracting a digital template from the data, and are more
affected by galaxy surface brightness than galaxy morphology. Additional
complications arise because magnitude-limited selection might prevent extinguished objects from being
discovered with equal efficiency in both nearby and distant samples, or because the smaller
angular size of distant objects makes it more difficult to detect SN deep inside spirals --- those likely to be
extinguished. In short, it would not be
surprising to have systematic differences between the
average extinctions of objects discovered in the nearby and distant supernova
searches.  Rather than attempt to untangle this hopelessly
knotted set of selection effects, we believe it is more straightforward to
correct for extinction of individual objects that are in the samples based on their colors.

Except for a small group of very rapidly fading supernovae, H95 show that SN~Ia have a 
relatively small range in their $(B-V)$ color at maximum light which can be used to 
estimate their extinction. Even so, it is worthwhile to correct for extinction
according to the prescription of P93/H95 or RPK95.  Using the framework set out
earlier in this section, but for two filters, a relative reddening estimate of
any SN~Ia can be extracted from its observations, by simultaneously
fitting $B$ and $V$ (or any set of two or more photometric bands) and
deriving the $E(B-V)$ offset. This procedure does not
require the training set to be free of reddening, as the
calculated color excess will be relative to the average extinction of
the training set. In practice, using an iterative process,
it is possible to define an essentially reddening-free set of 
synthetic light curves, weeding out the reddened objects  on the basis
of their color evolution.  RPK96 have
demonstrated that using $B,V,R,$ and $I$ light curves within their framework,
dispersions of less than 0.15~mag are attainable. Furthermore, they show
that their best estimate of the reddening law from a few heavily reddened objects in
their sample of SN~Ia is indistinguishable from the mean Galactic reddening law
(Riess, Press, \& Kirshner 1996b \markcite{RPK96b}).
A recent   study by Phillips et al. (1998)\markcite{P98} based on previous work by Lira (1996)\markcite{L96}
on the intrinsic color evolution of SN Ia has also developed a consistent 
method to deredden type Ia light curves.
Correcting SN~Ia for extinction
does not necessarily decrease their dispersion as distance indicators (extinction corrections only
decrease the scatter when the error in the derived extinction is less than the
dispersion caused by extinction to a sample of objects), but is essential to
remove the systematic bias that absorption might introduce. 

It is possible that the average properties of dust might evolve as a function of look-back time, e.g., 
the LMC and SMC have extinction laws which differ significantly from that of the Milky Way (Bouchet et al. 1985\markcite{Bouchet85}).
Such an effect could significantly bias the comparison of nearby and distant objects 
if the average extinction to either the nearby or distant sample is large. Although the nearby sample of
objects has few reddened objects (RPK96), and indications are that the
high-z sample has low extinction as well (Riess et al. 1998), this possible bias warrants further investigation.
For example, observations of reddened objects in rest-frame $BVR_CI_C$ at large redshifts could be used to investigate
the extinction law at these look-back times, and limit uncertainties introduced by this effect.

	\subsection{K-corrections}

When comparing objects at significant redshifts with nearby counterparts, it is necessary
to account for the wavelength shift of light.  The effect of redshift on luminosity
distances, dubbed the K-correction, was tabulated for galaxies in modern photometric bands by
Oke \& Sandage (\markcite {OS68}1968).  They defined the K-correction $K_i$ such that an object's
magnitude $m_i$ in filter $i$ as a function of redshift $z$ is
\begin{equation}
m_i(z) = m_i(z=0) + K_i(z).\label{eq: Kcorrdef}
\end{equation}
An object with spectrum $F(\lambda)$ (in units of power per unit area per unit
wavelength) observed in a filter with  sensitivity function $S_i(\lambda)$ has
\begin{equation}
m_i(z=0) = -2.5\log{{\int S_i(\lambda) F(\lambda)d\lambda} \over
{\int S_i(\lambda) d\lambda} } + {\cal Z}_i, \label{eq:magz=0}
\end{equation}
where ${\cal Z}_i$ is the zero point of the filter\footnote[18]{Traditionally
${\cal Z}_i$ has been set such that  when equation (\ref{eq:magz=0}) is applied
to the spectrum of Vega, it yields $m_i=0.03$~mag. However, many photometric
systems have ${\cal Z}_i$ which vary slightly from this definition.}.
Oke \& Sandage demonstrated that
\begin{equation}K_i = 2.5\log\left[ (1+z) {\int F(\lambda) S_i(\lambda)d\lambda \over
\int F(\lambda/(1+z))S_i(\lambda)d\lambda}\right],\label{eq:Kcorr}
\end{equation} 
where $K_i$ accounts for the $(1+z)$ shift
of the photons in wavelength, and the $(1+z)$ increase in the unit
$d\lambda$ which they occupy. In the case of SN~Ia, K-corrections have an
added complication caused by the changing spectral energy distribution of a
supernova as it evolves. SN~Ia change rapidly when they
are near maximum, which translates to rapidly changing (and hence
uncertain) K-corrections (Hamuy et al. 1993\markcite{H93b}). This evolution is particularly pronounced
blueward of 4000~\AA~ due to line blanketing, and therefore using K-corrections
in the traditional manner on the observed $B$ and $V$ light curves of SN~Ia at $z=0.5$
is an extremely risky proposition.  Following the precepts of Gunn (\markcite{G78}1978),
Perlmutter et al. (\markcite{Pe95}1995) demonstrated that by observing distant SN~Ia
with the $R_C$ broadband filter, K-correction uncertainties could
be substantially reduced, because at $z=0.5$ the $R_C$ filter
approximates the restframe $B$ filter. Kim, Goobar, \& Perlmutter (\markcite{K96}1996)
show that this modified K-correction is 
 
\begin{equation}K_{ij} = 2.5\log\left[(1+z) {\int F(\lambda) S_i(\lambda)d\lambda \over
\int F(\lambda/(1+z))S_j(\lambda)d\lambda}\right] + {\cal Z}_j - {\cal Z}_i, \label{eq:newKcorr}
\end{equation}
where $K_{ij}(z)$ is the correction for going from filter $i$ to filter $j$.  Note that if
filter $i$ is exactly the redshifted counterpart of filter $j$, then at this
redshift $K_{ij}$ is not zero, but has a constant value corresponding to the
$(1+z)$ stretching of the wavelength region $d\lambda$ photons occupy
combined with the difference in the zero points of the filters.

Rather than constrain ourselves to existing broadband filter sets, we
decided to define our own set of redshifted filters, created as 
broadband interference filters. We currently have defined four filters ---
pairs of $B$ and $V$ filters redshifted to $z=0.35$ and $z=0.45$ (henceforth
referred to as the $B35$, $V35$, $B45$, and $V45$ filters). Although adopting this
new photometric system adds the complication of defining standards for these
filters, we believe the ability to translate
SN~Ia observed in the range $0.25 < z <  0.55$ to standard $B$ and
$V$ filters with small systematic and statistical uncertainties 
makes it worthwhile. An overview of our system is given in Appendix A.

An important advantage in designing new filters is that modern transmission
coatings have peak transmissions in excess of $85\%$, whereas traditional $R_C$ and $I_C$ filters
are typically around $60\%$. In addition, our filters have a  sharp cut-off on the
long wavelength side, which results in significantly lower backgrounds from night sky lines than their
Kron-Cousins counterparts.
In Figure \ref{fig:Kcorrsig} we have plotted the uncertainty in the K-corrections in translating
to restframe $B$ and $V$ from both our specialized filters (Appendix A), the $R_C$ and $I_C$
filters (Bessell \markcite{B90}1990), and selected {\it HST} filters. These corrections have been calculated
using the series of SN~Ia spectra in Hamuy et al. \markcite{Hetal93}(1993b), augmented with
additional spectra of SN~1994D by Filippenko \markcite{F97}(1997). The resulting K-corrections
are fit as a function of SN age for each redshift with a low-order polynomial, and
the residual scatter of this fit is used to gauge the uncertainty of the K-corrections.
The estimated uncertainties are insensitive to the exact order of the polynomial used, and
provide an estimate of the random component of K-correction uncertainties due to errors 
in our SN~Ia spectrophotometry and intrinsic differences in supernovae.
For proper redshift-filter matches, the resulting K-corrections in transforming to restframe
$B$ and $V$ are accurate to  better than 3\% for SN~Ia observed between 14 days before maximum
and 50 days past maximum over the range of $0.25 < z < 0.55$ (and using {\it HST}, between $0.90<z<1.1$). 
For objects which are extinguished, it is necessary to modify the spectra used
for determining the K-corrections iteratively with reddening curves, using fits to their light curves
to estimate the amount of extinction. In addition, if the SN differs significantly 
from those used to construct the K-corrections, a substantial error can result, especially when the object 
is at a redshift where the filter is not a close match to its restframe counterpart.
In these cases, the first order difference between the spectra is usually color, and the spectra 
used for determining the K-corrections can be adjusted with a power law to match the color evolution
of the object. The systematic uncertainty in translating between filter systems limits the accuracy
with which we are able to measure luminosity distances at high redshift. Note that 
K-correction zero point errors are magnified through extinction corrections: since $A_V=3.1 E(B-V)$,
a $0.02$~mag uncertainty of $(B-V)$ translates to a 0.06~mag uncertainty in distance modulus. It is a
very difficult task to ensure  systematic errors in $(B-V) < 0.02$~mag in either the nearby or distant sample,
and we believe that this uncertainty will be our largest source of systematic error.

	\subsection{Evolution}

An attractive feature of using SN~Ia as distance indicators at significant redshifts
is the possibility of minimizing the evolutionary effects which plague
distance indicators based on the properties of galaxies.  Initial models
proposed that SN~Ia arise from the explosions of carbon-oxygen white dwarfs as they
reach the Chandrasekhar mass (\markcite{HF60}Hoyle \& Fowler 1960; \markcite{A69}Arnett 1969;\markcite{CM69} Colgate \& McKee 1969).
This mechanism for SN~Ia explosions leads to a burning front which propagates
outwards from the white dwarf's center, burning nearly the entire star to nuclear statistical
equilibrium. A broader range of models for the presupernova star and for the behavior
of the burning wave is needed to account for the intermediate-mass elements that are seen
in the spectra of SN~Ia (\markcite{WH90}Wheeler \& Harkness 1990), and to reproduce the observed
range of SN~Ia luminosities. Successful models have been produced by several
groups employing a variety of mechanisms (\markcite{NTY84}Nomoto, Thielemann,
\& Yokoi 1984; \markcite{L90}Livne 1990; \markcite{K93}Khokhlov, M\"uller, \& H\"oflich 1993;
\markcite{WW94}Woosley \& Weaver 1994; \markcite{HKW95}H\"oflich, Khokhlov, \& Wheeler 1995). Although
the details of the explosion mechanism remain an active area for research, many plausible (but similar) models match the spectral features of SN~Ia, and can produce the relation between light curve shape and luminosity detected by P93.

The local sample of SN~Ia shows that the light curve shapes and luminosities are weakly correlated
with the type of galaxy in which they occur (H96a). Spirals show a wide range of light curve shapes
while ellipticals show a narrow range. If left untreated, the relation between stellar population and 
the luminosities of SN~Ia could poison the inference of cosmological parameters by introducing a subtle
drift in SN~Ia properties with look-back time (von Hippel, Bothun, \& Schommer 1997). We know that the
rate of SN~Ia per unit $B$ luminosity is almost twice as high in spirals as in ellipticals at
the present epoch (Cappellaro et al. 1997)\markcite{C97}. If we assume this increased rate of SN~Ia production
is related to the higher rate of star formation in spirals, then the SN~Ia in spirals come from
progenitors which are likely to be younger than the progenitors of SN~Ia in ellipticals. Incidentally,
this is the conclusion reached by Oemler and Tinsley (1979) nearly two decades ago. 

However, our local sample calibrates the effect. Figure \ref{fig:evol} shows that the calibration
of RPK96, which does not use any information about galaxy type as an input, results in distances
to early-type (8 objects) and late-type galaxies (19 objects) which are consistent to $0.006 
\pm 0.07$ magnitudes (SN~Ia with elliptical hosts give slightly closer distances). To obtain limits
on the possible offsets between SN~Ia with early-type and late-type progenitors, we perform a Monte Carlo
simulation with the following assumptions. Late-type galaxies have SN~Ia with young and old 
progenitors at a ratio of 1:1 (a conservative limit from their rates  --- sprials have twice the
rate of supernovae as ellipticals, so 1/2 of the progenitors are old and 1/2 are young), and ellipticals contain
only old progenitors.  Simulated data sets demonstrate that with the above assumptions, and the observed
offsets between distances to ellipticals and spirals, the allowed range in the distance
offsets between objects with young and old progenitors is $0.02 \pm 0.15$ mag. If the ratio of young
to old progenitors in spirals were to be 2:1, then the limits become  $0.01 \pm 0.11$ mag.
This unsophisticated analysis, while not providing hard limits on evolution, demonstrates there is
no obvious dependance on SN~Ia distances with respect to the age of the stellar population
in which they reside. Since, at  $z=0.5$, we still expect to see SN~Ia originating from a mixture
of young and old progenitors, the average evolution should be smaller than the maximum possible differences
quoted above.
A larger sample of objects and a better understanding of the local rates as
a function of progenitor type, as will be provided by the Mount Stromlo Abell Cluster
Supernova Search (\markcite{Reiss98}Reiss et al. 1998), will increase the power of this type of test.

Theory provides another avenue by which to explore the possible effects of evolution. H\"oflich, Thielemann \& Wheeler(1997)\markcite{HTW97} have calculate the differences in light curve shape, luminosity,
and spectral characteristics of SN~Ia as a function of the initial composition and metallicity of their
white dwarf progenitors. Their calculations show that although changes in restframe $U$ can be considerable,
the effects in the  $B$ and $V$ bands are only $\sim 0.05$ mags, even when considerable changes are made
to the metallicity and C/O ratio of the white dwarf progenitors. Changes to the light curve shapes of the
objects may also occur, and in total their calculations suggest distances could drift by as much as 0.3~mag
in $B$ and 0.15~mag in $V$.  At what redshift such differences could appear is unclear, but these types of
changes would be accompanied by significant spectral differences, and should not go undetected.

We assume that the relation between light curve shape and luminosity that holds for a wide range of stellar
populations at low redshift also covers the range of stellar populations we encounter in our high-redshift
sample. The range of ages for SN Ia progenitors in the nearby sample is likely to be larger than the look-back
time to the galaxies in our high-redshift sample, so we expect that our local calibration will
work well at eliminating any pernicious drift in the supernova distances between the local and distant
samples.  Although we expect this approach to be valid for joining our nearby and distant samples, until
we know more about the stellar ancestors of SN~Ia, we need to be vigilant for changes in the properties
of the SN at significant look-back times. These might be detected as changes in the spectra or colors
of SN Ia with redshift. 

	\subsection{Selection Biases}

Like every sample selected by a flux limit, our SN~Ia sample will be affected by the
shift of the mean to intrinsically brighter objects near our redshift limit due to the 
dispersion of their intrinsic luminosities; this is often referred to as Malmquist bias
(e.g., Gonzalez \& Faber 1997\markcite{GS97}).
Generally speaking, the bias is proportional to $\sigma^2$ of the distance
indicator.  Since we only suffer the bias left {\it after} the correction for light
curve shape, the average  bias is much smaller than for uncorrected SN~Ia peak brightnesses.
This is fortunate, because it is not a straightforward task to correct for luminosity biases
in the limited sample of objects produced by a particular SN search.

In addition to classic Malmquist bias, the SN~Ia we discover at high redshift are subject
to a number of selection effects. These could affect the average properties of our
sample as a function of redshift, and could possibly bias our measurement of the ($D_L$,$z$) relation.
The supernovae discovered by the High-Z SN Search are limited by their
detectability between two epochs (typically 25 days apart in the observer's frame), and only
those objects which increase in flux by an amount greater than our detection threshold
are discovered. Therefore, our ability to discover a supernova depends not only on
its brightness at maximum light, but also on its age, the light curve shape of the object, and
redshift due to time dilation of the light curves.

We have performed  Monte Carlo simulations in which we randomly explode SN~Ia assuming their
rate is constant with look-back time, identifying which objects would be discovered in
a search by our supernova program. These simulations show that with SN~Ia 
distances of precision $\sigma=0.15$~mag, our average
measurement of $\Omega_M$ will be biased high by $\sim 0.05$ for $0 < \Omega_M < 1$ ( $\Omega_\Lambda=0$).
Though it may seem surprising, selection biases
are significantly less than other sources of error, such as the
uncertainty in $(B-V)$, because the scatter in luminosity, after correction for light curve shape,
is so small. Nevertheless, selection effects are a source of systemic error that should be removed.

Corrections could be computed through Monte Carlo simulations for equation (\ref{eq:intDL})
provided that the selection function (completeness vs. magnitude) for finding 
objects were known.  Unfortunately, the selection function for discovering SN~Ia is not sharply defined;
it depends critically on the PSF of both the pre-discovery and
discovery images, and our procedure ultimately rests on a human decision of whether
an object is real. Since false detections waste scarce observing time,
observers are reluctant to follow the light
curve of an object unless the probability the object is real is very high.
Perlmutter et al. \markcite{Pe97}(1997) performed simulations
to determine the magnitude limit for each of their discovery frames, adding stars
to galaxies and then measuring their ability to recover them in software. Our experience
with similar simulations is that our software selects fainter objects than
human reviewers will accept as being real. By comparing Monte Carlo simulations of the
expected redshift distribution of objects to the actual sample, it should be possible
to estimate corrections to these detection thresholds.

Rather than apply corrections to individual objects, it would be better
to employ a maximum likelihood approach in which one finds the most likely cosmological
model, given the SN redshifts, estimated detection thresholds, SN distances and light curves,
and detection simulations. This is an involved, computationally expensive procedure,
which is probably not yet warranted given the small size of the corrections and the limited
set of data.

\subsection{Weak Gravitational Lensing}
	
	It has been pointed out by Kantowski, Vaughan, \& Branch (\markcite{KVB95}1995) that large-scale
structure could magnify (or demagnify) a SN's light through weak gravitational
lensing as it travels to an observer.  Wambsganss et al. (\markcite{W97}1997) have computed the
effect for $\Omega_\Lambda$-dominated
and $\Omega_M$-dominated  flat universes, and find that weak lensing could produce a modest increase in the
dispersion of a distance indicator (approximately 5\% at $z=1$). It also leads to a small systematic
shift in observed SN brightnesses to fainter magnitudes --- approximately 1\% at $z=1$.  The average
line-of-sight to a SN is more likely to pass through voids than clusters and filaments, leading to an
average demagnification.  Holz and Wald (1997)\markcite{HW97} have calculated a ``worst case'' scenario for
the effect of gravitational lensing, where the universe's mass is made up of randomly distributed objects
with mass greater than $0.01 M_\odot$. At $z=0.5$ for $\Omega_M=1$, an average standard candle is
made 0.15~mag fainter by the lensing. The effect is more pronounced at $z\approx1$, and is diminished
as $\Omega_M$ approaches 0, as suggested by the SN observations presented here, by Garnavich et al. (1998),
and by Perlmutter et al. (1998).
Given the size of other systematic errors, uncertainties due to gravitational lensing are not likely to be of
major concern up to $z\approx 1$. It is unlikely that we might find an event which has undergone significant
lensing so that it can be readily separated from the intrinsic dispersion of SN~Ia brightnesses, although
Kolatt \& Bartelmann (1998) \markcite{KB97}suggest how to maximize these events by searching through galaxy clusters.

\subsection{Summary of Uncertainties}

The data presented by Garnavich et al. (1998) and Riess et al. (1998) indicate that we can measure the
distances to high-z supernovae with a statistical uncertainty of $\sigma = 0.2$ mag (10\%)
per object. With only 10 objects a comparison of $z\approx0$ and $z\approx0.5$ can be made
to a precision of better than 5\% --- leaving systematic uncertainties as a major contributor
to the total error budget. A summary of the contributions to high-redshift supernova distance
uncertainties is given in  Table \ref{tab:sys}. This table shows  that 
our program to measure cosmology will  most likely be limited by the possibility
of the evolution of SN~Ia explosions with look-back time. Future work to address this
possible problem will be as important as obtaining large numbers of objects at high redshift.

\section{ Observations and Analysis of SN 1995K}

\subsection{ Photometry and Spectroscopy}

In 1995 February and March our team imaged approximately 2.5 square degrees with the CTIO 4~m telescope
+ 2048$^2$ CCD. Our first attempts at searching for supernovae in these data were hampered by distortions caused
by curvature in the CCD, and general inexperience. However, in an image taken on 30 March (UT dates are used throughout this paper),
a new object was discovered in a galaxy at position (J2000.0) $\alpha=10^h50^m47^s.0$,
$\delta=-09^\circ15^\prime07^{\prime\prime}.4$ (Figure \ref{fig:95Kbanda}). This object was
reported to the IAU and designated SN 1995K (Schmidt et al. 1995\markcite{S95}).

We obtained spectra of SN~1995K using the ESO~NTT+EMMI on 1995 April 3. CCD images
containing the SN spectra were bias subtracted, and the spectrum of the object extracted
as a single pixel row, to minimize the contribution of the host galaxy. The spectra were then
wavelength calibrated with  comparison lamp spectra interspersed with the SN observations,
and the resulting individual spectra were combined. The extracted host galaxy spectrum
was scaled and subtracted.  The host galaxy's redshift, measured from its H$\alpha$ emission, is
$z=0.479$. The spectrum was binned to increase the S/N ratio per
resolution element, and the resulting spectrum is plotted in Figure \ref{fig:95Kspec}.  Also plotted
is the spectrum of SN~1994D (\markcite{F97}Filippenko 1997) near maximum light, binned to the
same resolution as SN 1995K. The spectrum of SN~1995K appears to be typical of
SN~Ia, exhibiting  the characteristic Si~II absorption at a rest wavelength of $\lambda6150$~\AA~ and
is incompatible with peculiar SN~Ia such as SN~1991T (\markcite{F92}Filippenko et al. 1992a;
\markcite{P92}Phillips et al. 1992; \markcite{RP92}Ruiz-Lapuente et al. 1992; \markcite {Ford93} Ford et al. 1993) or SN~1991bg 
(\markcite{F92b}Filippenko et al. 1992b; \markcite{L93}Leibundgut et al. 1993;  Turatto et al. 1996).
Using the spectral aging technique of Riess et al. \markcite{retal97}(1997), we estimate the
age of the SN on this date, 1995 April 3, to be $1\pm 2$ days past maximum light, {\it independent} of,  and in
agreement with its light curve. An additional spectrum of SN~1995K's host galaxy was obtained on 1995 Apr~25
with the CTIO~4~m telescope (Figure \ref{fig:95Kspec}). Comparison of the galaxy spectrum with the catalog
of Kennicutt \markcite{K92}(1992) shows this galaxy to be consistent with a star-forming galaxy of
type Sb/c. 

CCD images of SN~1995K were obtained at several telescopes in $B45$ and $V45$ filters, and in some cases
$R_C$ and $I_C$. The sequence of stars shown in Figure \ref{fig:95Kseq} was calibrated on three photometric
nights by observing spectrophotometric standards listed in Appendix A and standards of
[Landolt (\markcite{L92}1992a) for $R_C$ and $I_C$ measurements],  measuring the color and
extinction transformation coefficients, and applying these to observations of the SN~1995K field. The
resulting magnitudes were averaged from multiple nights and are given in Table \ref{tab:95Kseq}, with
uncertainties calculated from the dispersion of the observations.
Relative photometry between the supernova and the stellar sequence was
carried out as prescribed in \S4.3, using a template for galaxy subtraction obtained
with the CTIO~4~m+2048$^2$ CCD on 1996 March 15.  SN~1995K's light curve was brought to the standard
system by offsetting from the standard star sequence, and applying color corrections derived for
each instrumental setup (these corrections  ranged from $-0.01$ to $0.04$ mag per mag of
$(B_{\rm 45}-V_{\rm 45})$).  The resulting light curve is listed in Table \ref{tab:95Klc}. 

\subsection{The Luminosity Distance to SN~1995K}

The K-corrections to translate  $B45$ and $V45$ observations of SN~Ia at a redshift of $z=0.479$
to restframe $B$ and $V$, respectively, are  plotted in Figure \ref{fig:95KKcorr}. The K-corrections
are nearly constant, because of the close match of these redshifted filters to their restframe
counterparts. Table \ref{tab:95Ktab} lists each corrected photometric measurement, the time-dilated
age of the SN with respect to maximum fit from the light curve (iteratively
determined by applying K-corrections, fitting the light curve, and refitting the K-corrections, until
convergence occurs), and the resulting restframe magnitudes for SN~1995K.  The light curve of SN~1995K
 has already successfully been used to provide strong evidence for the predicted effects of time dilation
(\markcite{L96}Leibundgut et al. 1996) in SN light curves (\markcite{L90}Leibundgut 1990).

The light curve of SN~1995K is plotted in Figure \ref{fig:95Klc}, and is of sufficient quality to
provide an accurate luminosity distance to its $z=0.479$ host. Using the techniques described in
H95 we derive $\Delta m_{15}(B) = 1.15\pm 0.1$~mag for SN 1995K, and  restframe brightnesses at maximum 
light of $m_B=22.93\pm 0.08$~mag and $m_V=23.04 \pm 0.13$~mag. This decline rate and color at maximum are typical for the
unreddened SN~Ia we observe in the nearby Universe.
The derived distance modulus, applying the $\Delta m_{15}(B)$ versus $M_V$ relations of H96a (their Table~3), is
$(m-M)_V = 42.31 \pm 0.13$~mag, and the Hubble diagram
which includes SN 1995K and the H96b distance measurements to 29 objects is shown in Figure \ref{fig:95KhubH96}.
Applying MLCS (RPK96) to SN~1995K indicates that this SN is slightly over-luminous relative to the
average ($\Delta= 0.07$), is unreddened, and has a distance modulus of $42.40\pm0.25$~mag on the RPK96 scale.
This object and the distance
measurements of RPK96 are plotted in Figure \ref{fig:95KhubRPK96}. Distances derived to the nearby and distant
supernovae must be done in exactly the same manner, so the results of these  two ways of using the light curve 
shape are not plotted on the same diagram. However, one can compare the resulting ($D_L$, $z$)
relation, and the bottom halves of Figures \ref{fig:95KhubH96}
and \ref{fig:95KhubRPK96} demonstrate that a consistent result is obtained.

With a single object it is difficult to make serious conclusions about cosmological parameters,
regardless of the distance precision it offers, because there is no way to judge systematic errors in
an empirical way. However, taken at face value, if $\Omega_\Lambda=0$, SN~1995K yields an estimate for
the matter density of the Universe to be $\Omega_M = -0.2^{+1.0}_{-0.8}$, where we have taken the range to include
both the H96 and RPK96 distance uncertainties, extinction uncertainties, and uncertainties due the sources
of systematic error described in $\S$5. 
For a spatially flat universe composed of normal matter and a cosmological constant, we find
$\Omega_M = 0.4^{+0.5}_{-0.4}$, $\Omega_\Lambda = 0.6^{+0.4}_{-0.5}$. An $\Omega=\Omega_M=1$ universe is excluded with  greater than 80\% confidence from this single distance estimate.

Perlmutter et al. \markcite{\pe96}(1997) present luminosity distances to 7 high-redshift supernovae. Direct
comparison with our work is difficult, because not all of their objects were positively identified as 
SN~Ia (two of their objects have decline rates more extreme than any other known SN~Ia and do not
have confirming spectra), and their single-color results cannot be corrected for extinction 
in the host galaxy. However, taken on equal terms, SN 1995K suggests that $\Omega_M$ may be lower than
the central value of $\Omega_M= 0.88^{+0.69}_{-0.60}$ they find. Recent {\it HST} observations by the two
groups (Garnavich et al. 1998; Perlmutter et al. 1998) seem to confirm this view, yielding values of
$\Omega_M = -0.1 \pm 0.5$ and $\Omega_M = 0.2 \pm 0.4$, respectively. More data will elevate this discussion.

\section{The Future}
SN 1995K is the first of more than 30 confirmed SN~Ia discovered by the High-Z SN Search (Schmidt et al. 1995, 1997a, 1997b;
Kirshner et al. 1995; Suntzeff et al. 1996; Garnavich et al. 1996a,b, 1997a,b). The observations
presented here indicate that these objects, and those of Perlmutter et al. (\markcite{Pe97}1997),
should provide an accurate method of tracing out luminosity distances to high redshifts. Figure
\ref{fig:future} shows the level of precision likely to be achieved by a sample of $N$ SN~Ia with
observations of comparable precision to those of SN~1995K;
a measurement of $\Omega_M$ to $\pm 0.2$ should be achieved from objects at $z=0.5$ given our expected
systematic uncertainties. The High-Z SN Search  has been allocated sufficient orbits to follow about 8
SN~Ia with the {\it HST} in two photometric bands.  We have already obtained observations of three SN~Ia, one
of which lies at $z=0.97$ (Garnavich et al. 1998\markcite{G98}). As demonstrated in Figure \ref{fig:Kcorrsig}, this object lies in a window in which
it would be possible to obtain accurate restframe $B$ and $V$ observations of SN~Ia with {\it HST}. A sample of 10 such
$z=1$ objects, coupled with the $z=0.5$ sample, would provide enough information to separate
the individual effects of $\Omega_M$ and $\Omega_\Lambda$ (\markcite{G95}Goobar \& Perlmutter 1995). Figure \ref{fig:future1} shows the $1\sigma$ joint
confidence region for $\Omega_M$ and $\Omega_\Lambda$ that is obtainable with a sample of 30 SN~Ia observed
at $z=0.5$ augmented with  10 SN~Ia observed at $z=1$ with {\it HST}. As can be seen from the figure, although $\Omega_M$ and
$\Omega_\Lambda$ are not individually measured accurately, the combination $\Omega_M - \Omega_\Lambda$ is
tightly constrained. If we limit ourselves to flat cosmological models, $\Omega_M +\Omega_\Lambda=1$, the
uncertainty in $\Omega_M$ and $\Omega_\Lambda$ would become small, with uncertainties in each quantity 
of approximately $0.1$.

Although measurements of the CMB anisotropy on small scales still have several years
to reach maturity, the high-z SN~Ia and CMB observations are complementary. The error ellipses of these two
measurements in the $(\Omega_M$,$\Omega_\Lambda)$ plane will be nearly perpendicular to each other (Figure \ref{fig:future1}).
When combined, they could provide a definitive measurement of the global parameters of our Universe.
\vskip 20pt
We wish to thank Ed Carter at NOAO for tracing the redshifted filter set. AVF acknowledges support from
NSF grant AST-9417213. RPK acknowledges support from NSF grants AST-9528899 and AST-9617058, and thanks the
Institute for Theoretical Physics, University of California, Santa Barbara for their generous hospitality.
AGR acknowledges support from the Miller Institute for Basic Research in Science, University of California,
Berkeley. SN research at UW is supported by the NSF and NASA. CWS acknowledges the generous support of the Seaver
Institute and the Packard Foundation. MH acknowledges support provided for this work by theNSF through grant number GF-1002-97 from the Association
of Universities for Research in Astronomy, Inc., under NSF Cooperative
Agreement No. AST-8947990 and from Fundaci\'{o}n Andes under project C-12984;
MH also acknowledges support by C\'{a}tedra Presidencial de Ciencias 1996-1997.
Partial support for AC was provided by the NSF through grant GF-1001-95 from
AURA, Inc., under NSF cooperative agreement AST-8947990, and from
Fundaci\'on Antorchas Argentina under project A-13313. This research used IRAF, an astronomical
reduction package distributed by the National Optical Astronomy Observatories, which is operated by 
the Association of Universities for Research in Astronomy, Inc. (AURA) under cooperative agreement
with the NSF.

\vfill
\eject

\appendix 

\section {Appendix: The High-Z Standard System}

We have created a new set of broadband filters which represent the Johnson $B$ and $V$ filters redshifted
to $z=0.35$ and $z=0.45$. We call these $B35$, $V35$, $B45$, and $V45$, respectively.
Traces of each filter have been supplied by the manufacturer, {\it Omega Optical}, and these traces have been verified
at KPNO using a spectrophotometer. The sensitivity functions $S_{B35}$, $S_{V35}$, $S_{B45}$, and $S_{V45}$
of our broadband filters have been derived by combining these filter traces with a 
quantum efficiency curve of a thinned SITE CCD and a normalized function which increases linearly with
$\lambda$ (Table \ref{tab:sens}). This latter function is included because CCDs are photon
detectors and standard sensitivity functions (e.g., \markcite {B90}Bessell 1990) assume
that a star with spectrum $F_{\lambda}$ in filter $i$ has magnitude $m_i$  given by
\begin{equation}
m_i = -2.5\log{{\int S_i(\lambda) F_\lambda(\lambda)d\lambda} \over
{\int S_i(\lambda) d\lambda} } + {\cal Z}_i. \label{eq:mag}
\end{equation}
For the purposes of the High-Z SN Search, the exact definition of our zero points ${\cal Z}_i$ is not important.
As long as we use a consistent method in our calculations with equations (\ref{eq:newKcorr})
and (\ref{eq:magz=0}), our derived K-corrections are sensitive only to the difference in 
zero points between our redshifted filters and the Johnson filters, e.g. ${\cal Z}_{B} - {\cal Z}_{B45}$. Rather than establishing our standard system's zero point using Vega, we use the 
Hamuy et al. \markcite{H94}(1994) spectrophotometric standards. These stars have accurate
broadband photometry in the standard systems of Cousins \markcite{C80}(1980)
and Landolt \markcite{L92a}(1992b), the systems with which our nearby supernova observations
have been calibrated. This consistency is important because our program's goal is to make
as accurate a differential measurement as possible of SN brightnesses as function of redshift.

We define  ${\cal Z}_i$ of our broadband filters 
such that A-type stars with $(V-I)=0$~mag have $B35=V35=B45=V45 \equiv V$. 
Hamuy et al. (\markcite{H94}1994) have observed
a set of A-type stars (referred to as secondary standards) which have $V-I_C \approx 0$. 
Since the Hamuy stars do not have exactly $V-I_C = 0$, we must apply small corrections to
$m_V$ of the stars to obtain magnitudes in our filter system. We use equation (\ref{eq:mag}) to
derive the magnitudes of the tertiary stars of Hamuy et al. (\markcite{H94}1994), which have $-0.3 < (V-I_C) < 0.8$, in our bandpasses, and then obtain the linear transformation between our system
and the Johnson/Kron-Cousins systems of the form
\begin{equation}
m_i = a_{ij}(V-I_C) + m_j. \label{eq:trans}
\end{equation}

The transformations used to convert the Johnson/Kron-Cousins system to our new system
are listed in Table \ref{tab:stdsi} and the fits shown in Figure (\ref{fig:HZphot}).  We use these
values to transform the secondary standards of Hamuy to the high-redshift SN system, and these 
stars and their derived magnitudes are listed as the first 10 stars in Table \ref{tab:stds}.
Because of the near zero color of the Hamuy
secondary stars, the corrections, even for the $B35$ and $V35$ filters
which have large transformation coefficients, are very small ($ < 0.02$~mag).
We compute the ${\cal Z}_i$ for our filters by comparing the magnitudes for the Hamuy secondary stars
obtained from equation (\ref{eq:trans}) with the magnitudes obtained from equation (\ref{eq:mag}), and adjusting ${\cal Z}_i$ for each
filter so that the average offset between the two sets of magnitudes is zero. The resulting
${\cal Z}_i$ are given in Table \ref{tab:stdsi}. Table \ref{tab:stds} lists the $B35, V35, B45, V45$ magnitudes of the Hamuy tertiary stars 
using these ${\cal Z}_i$ with equation (\ref {eq:mag}). The Hamuy secondary and tertiary stars form
the primary standards for the high-redshift SN photometric system. In a future paper we will present observations of these standard stars and 
use them to calibrate selected Landolt (1992a) fields.

\newpage

\figcaption[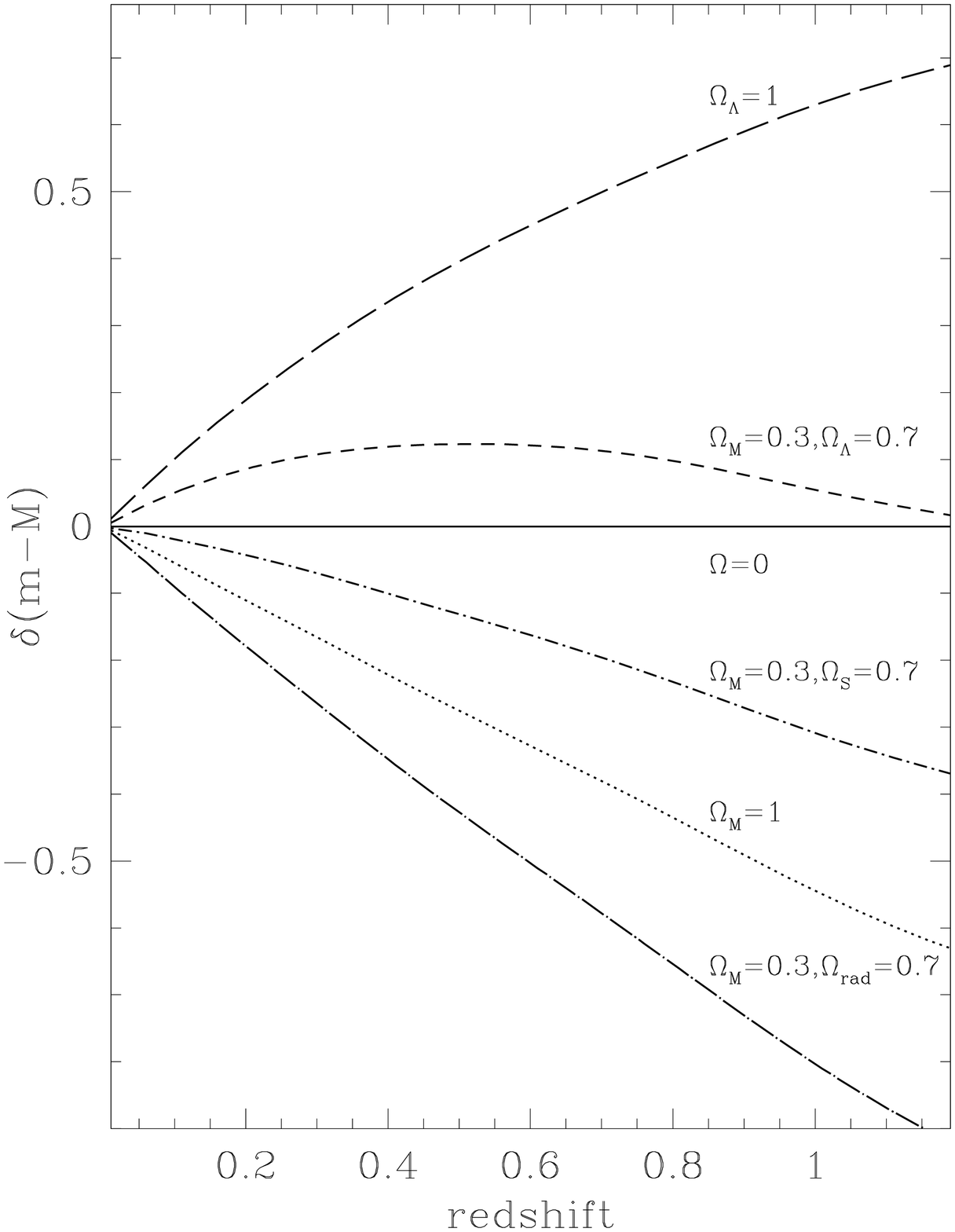]{The differences in the ($D_L$,$z$) relation for
various cosmological models expressed as the difference in distance modulus from an empty Universe,
$\Omega=0$.\label{fig:dlvsz}}

\figcaption[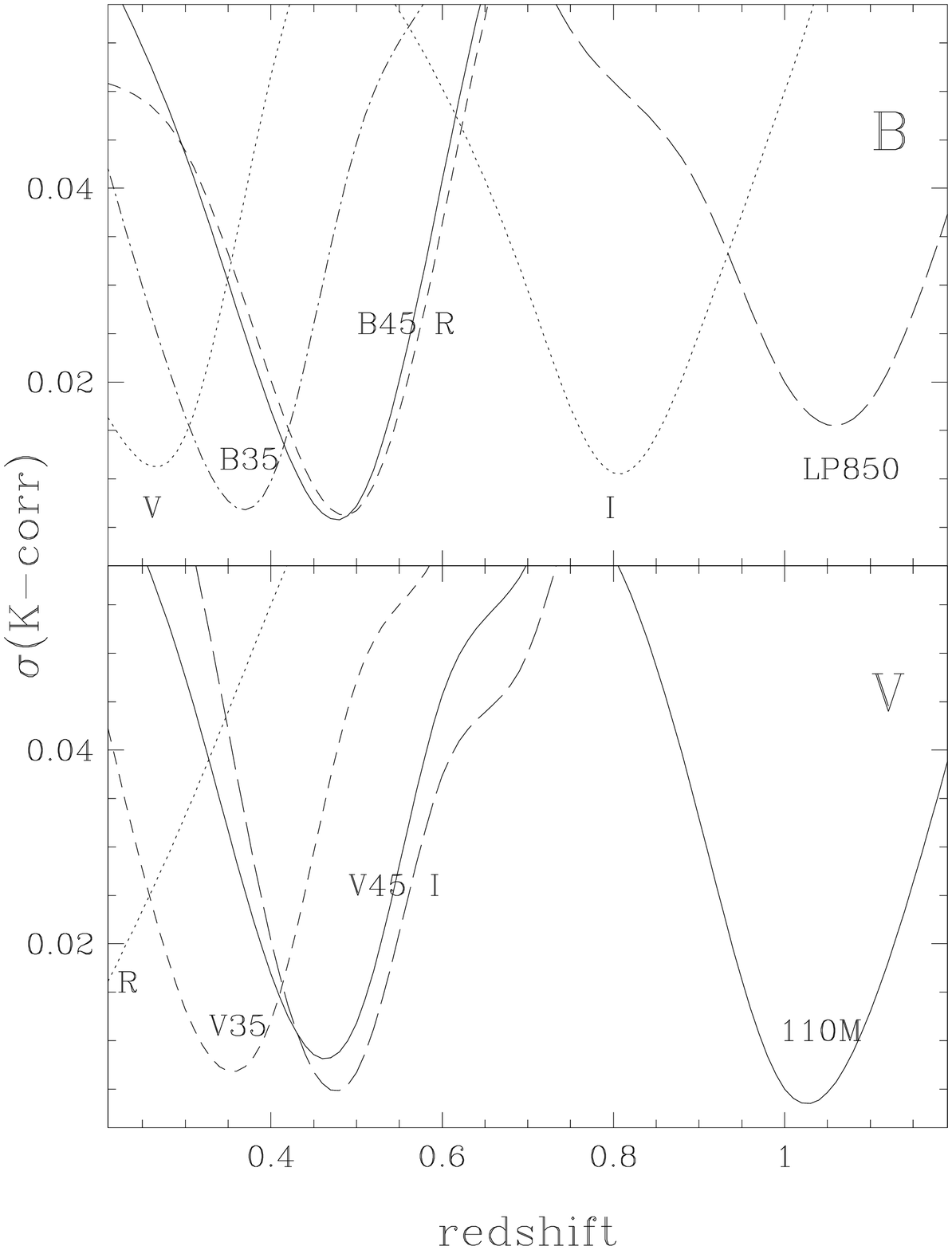]{The uncertainty in transforming to restframe $B$ (top) and $V$
(bottom) as a function of redshift for $V$, $R_C$, and $I_C$, {\it HST} 850LP (WFPC) and 110M (NICMOS),
and the $B35$, $V35$, $B45$, and $V45$ filters described in Appendix A. \label{fig:Kcorrsig}}

\figcaption[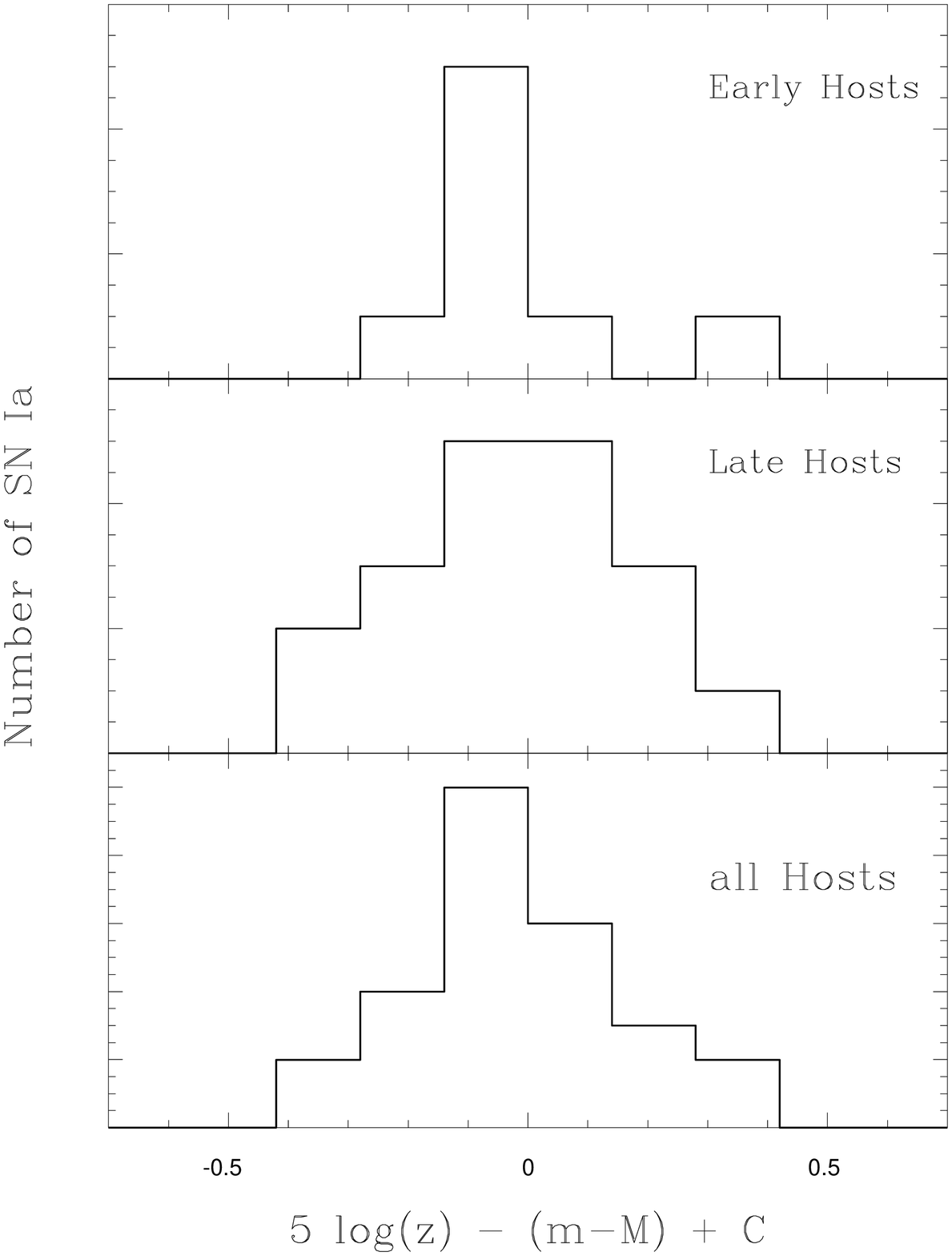]{The residual of SN~Ia distances from RPK96 plotted as a function
of galaxy type. The offset between the early-type and late-type galaxies is $0.006 \pm 0.07$~mag.
\label{fig:evol}}

\figcaption[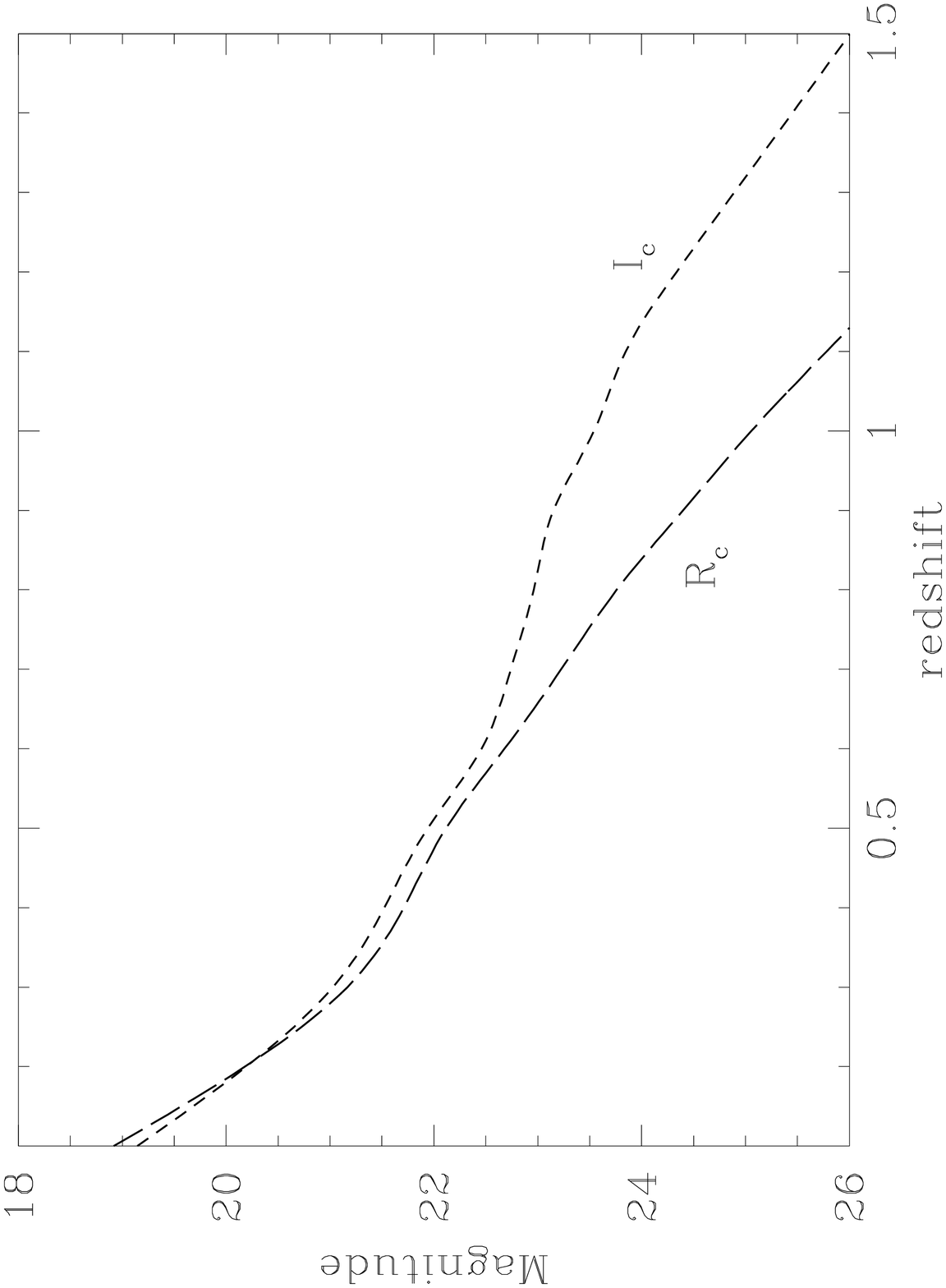]{The magnitude at maximum of a typical SN~Ia as a function of redshift for as observed in $R_C$ and $I_C$ for $\Omega_M=\Omega=0$. \label{fig:magvsz}}

\figcaption[fig5.ps]{The template image of SN~1995K taken 1995~March~7, the discovery image of
SN~1995K taken 1995 Mar 30, and the resulting subtracted image obtained as described in \S4.2.
\label{fig:95Kbanda}}

\figcaption[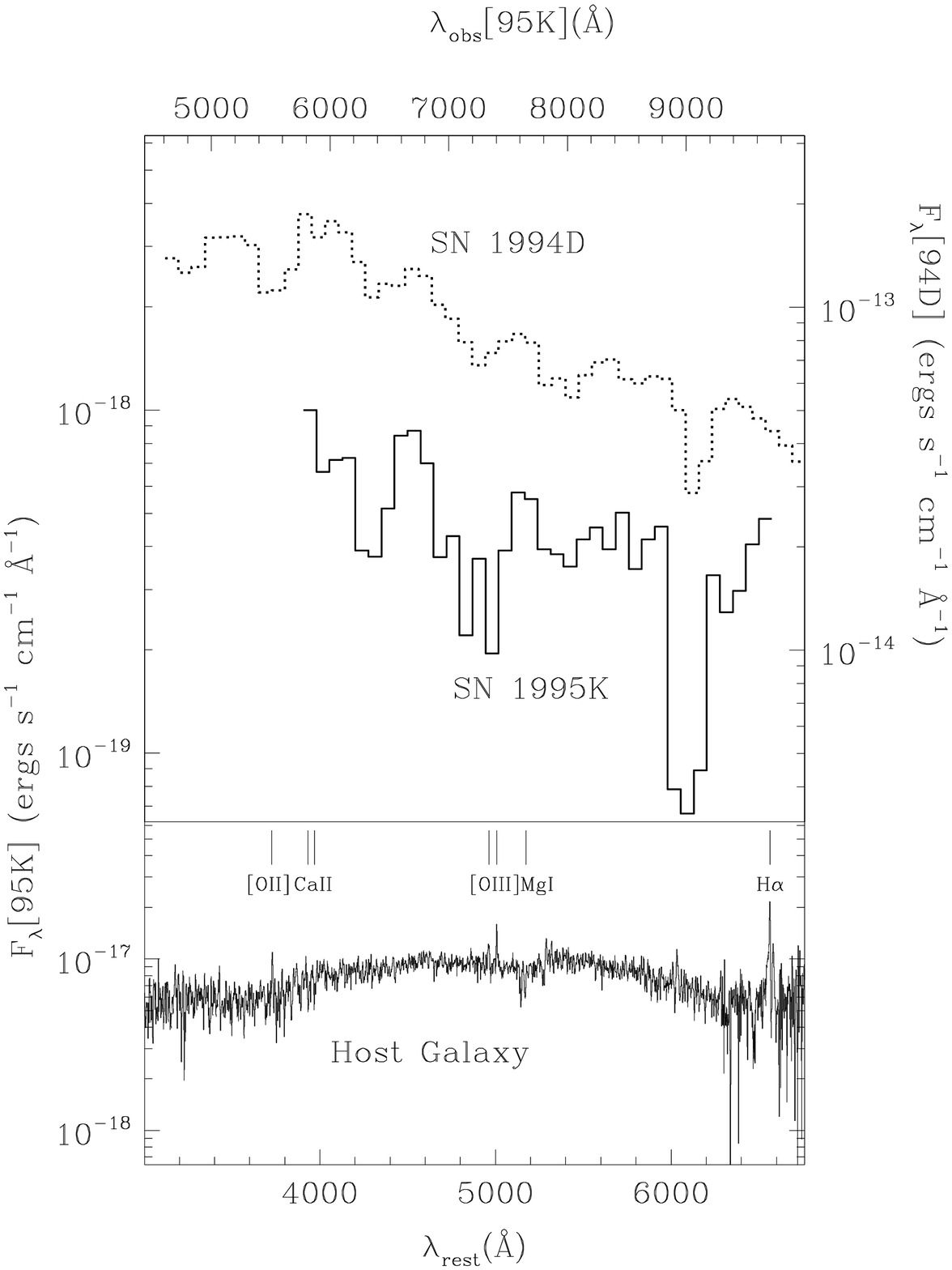]{ The spectrum of SN 1995K obtained with the ESO NTT on 1995 April 3 (UT),
compared to that of SN~Ia 1994D (Filippenko 1997) near maximum light. The
spectra have been binned to the same resolution, and the good match indicates SN~1995K is a SN~Ia caught near peak brightness. Also shown is the
spectrum of SN~1995K's host galaxy as obtained on 1995~Apr~25 with the CTIO~4~m
telescope.\label{fig:95Kspec}}

\figcaption[fig7.ps]{The field around SN~1995K as observed by the CTIO~4~m telescope, with the position
of the SN and photometric sequence stars marked.\label{fig:95Kseq}}

\figcaption[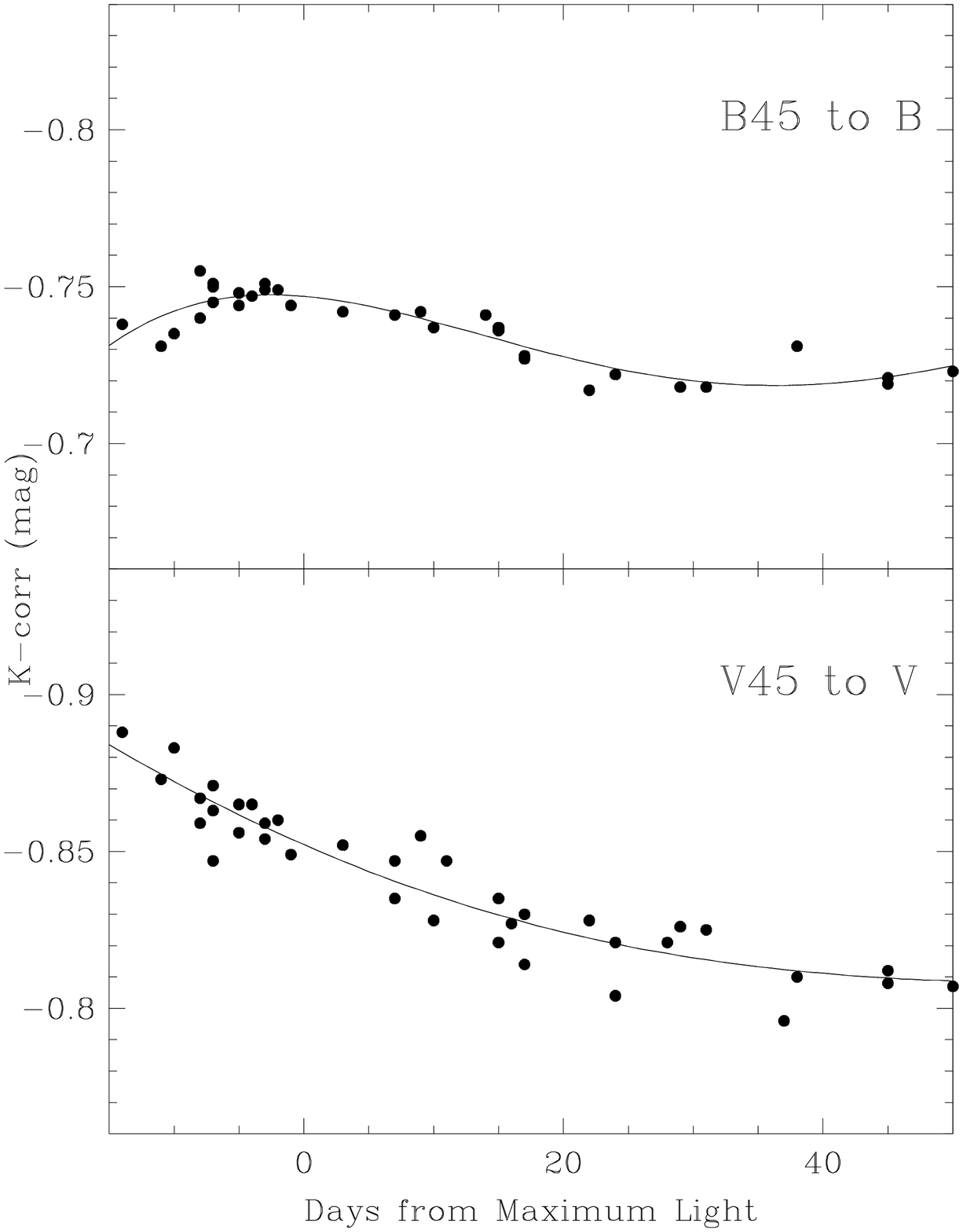]{The measured K-corrections for individual spectra plotted as a function
of supernova age for transforming $V45$ to $V$ (top) and $B45$ to $B$ (bottom) for SN~Ia at $z=0.479$. A
fifth-order polynomial is fit to the data.\label{fig:95KKcorr}}

\figcaption[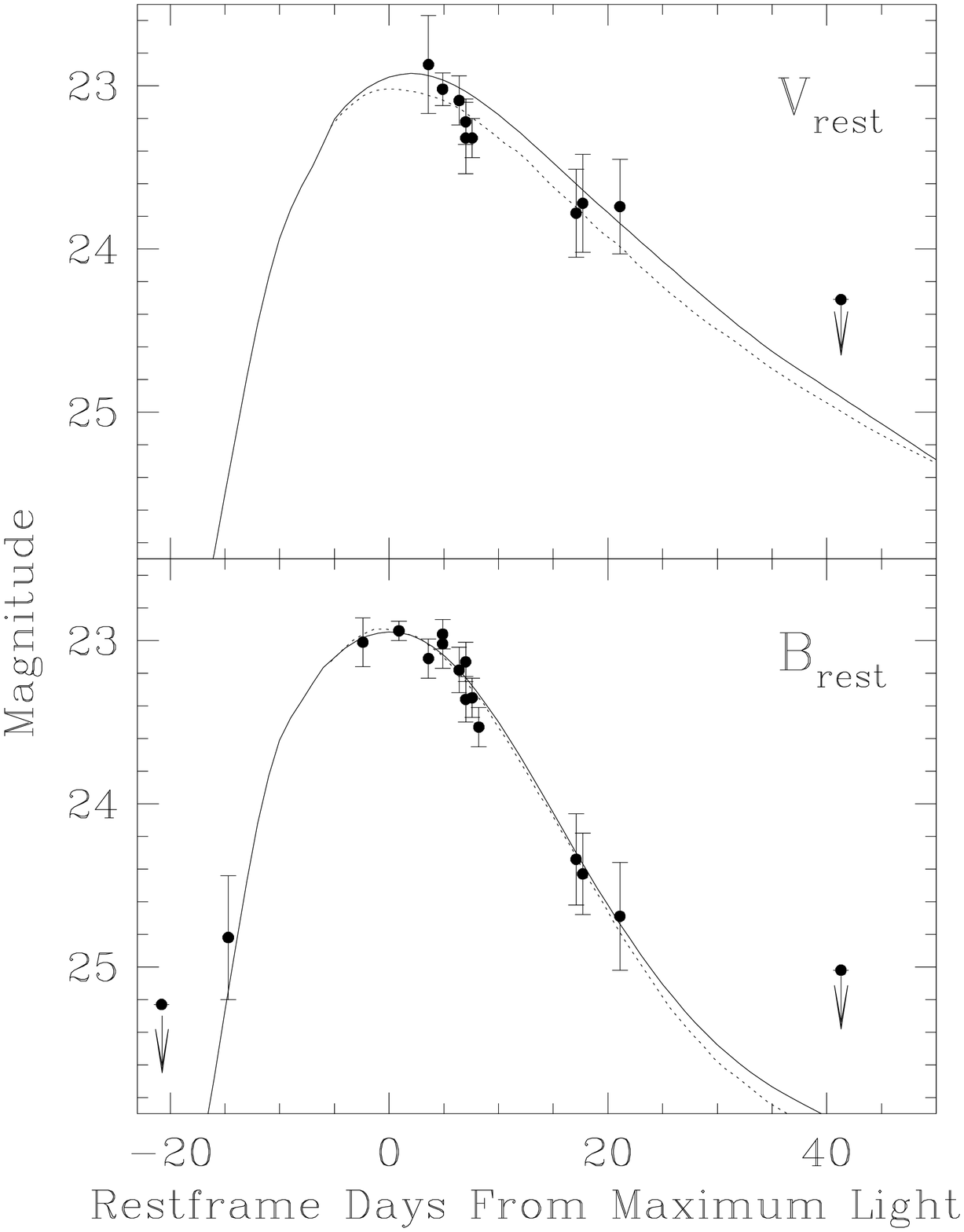]{The restframe $B$ (bottom) and $V$ (top) light curves of SN~1995K. The data are
K-corrected and time-dilated, and the RPK96 (solid line) and H95 (dotted line) fits are superposed. \label{fig:95Klc}}

\figcaption[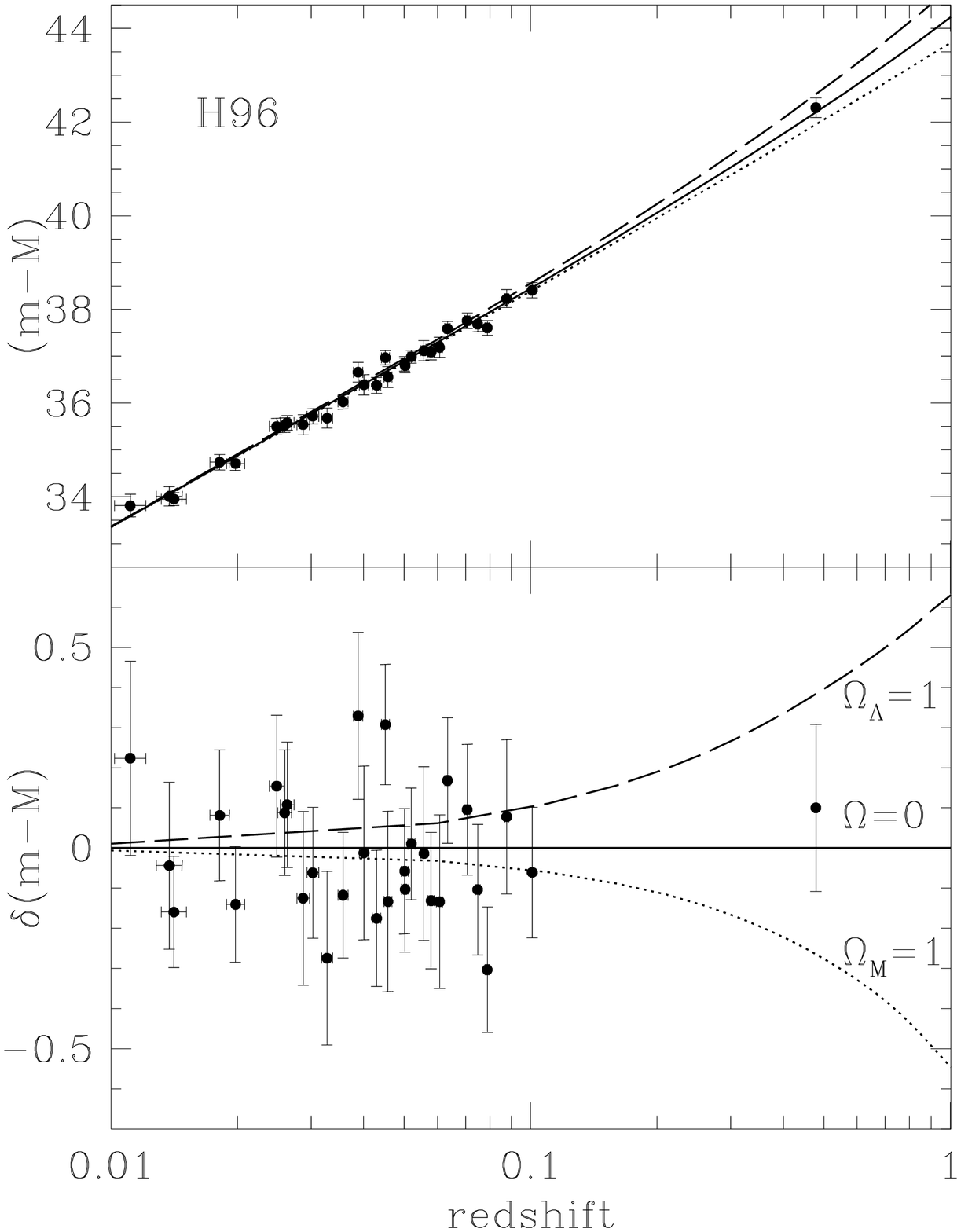]{A Hubble diagram containing SN~1995K using the distances and techniques of H96b(top), and these  distances plotted as a residual in distance modulus $(m-M)$, with respect to an empty ($\Omega=0$)
Universe. \label{fig:95KhubH96}} 

\figcaption[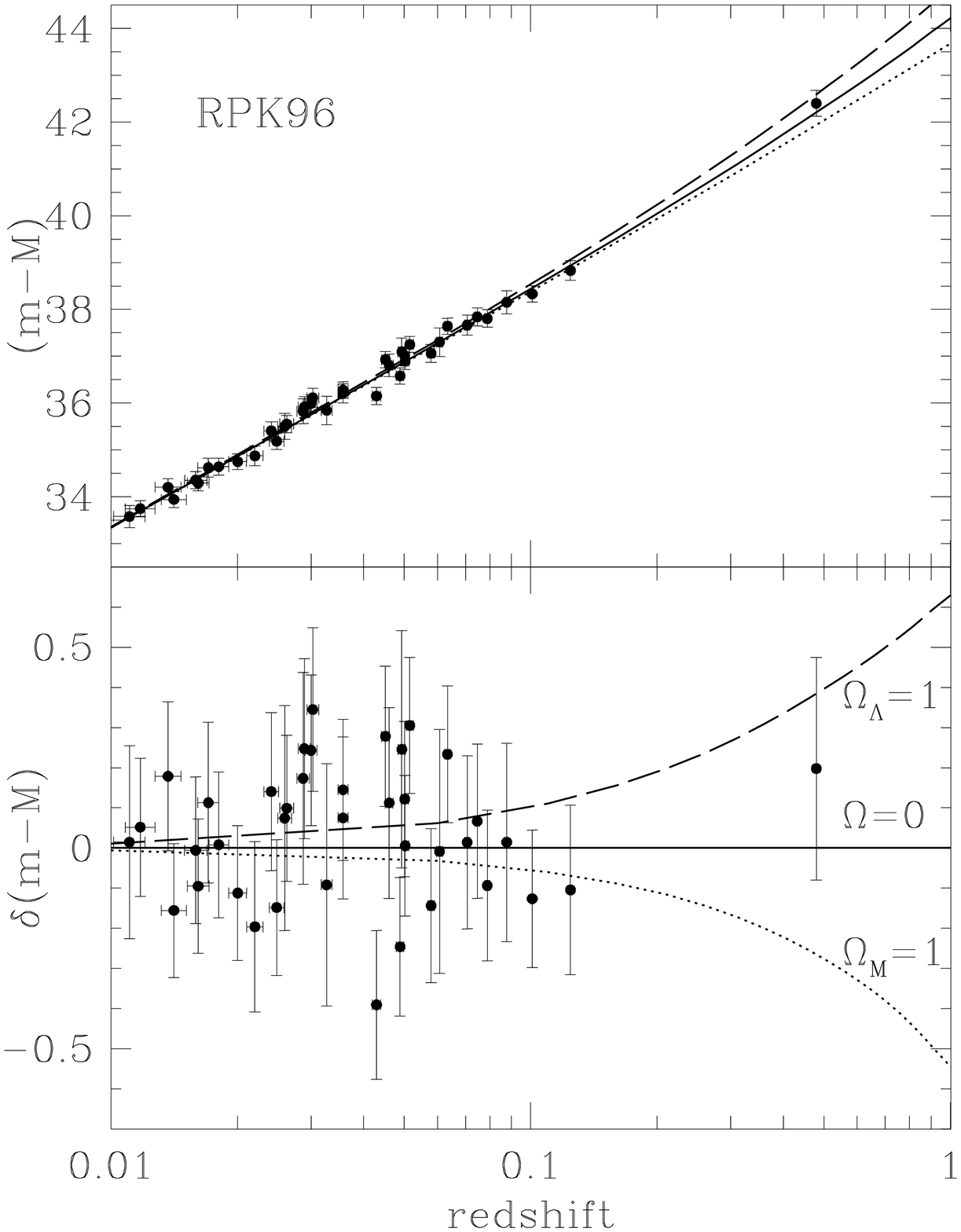]{A Hubble Diagram containing SN~1995K using the distances and techniques of RPK96 (top), and these distances plotted as a residual in distance modulus $(m-M)$, with respect to an empty ($\Omega=0$)
universe. \label{fig:95KhubRPK96}} 

\figcaption[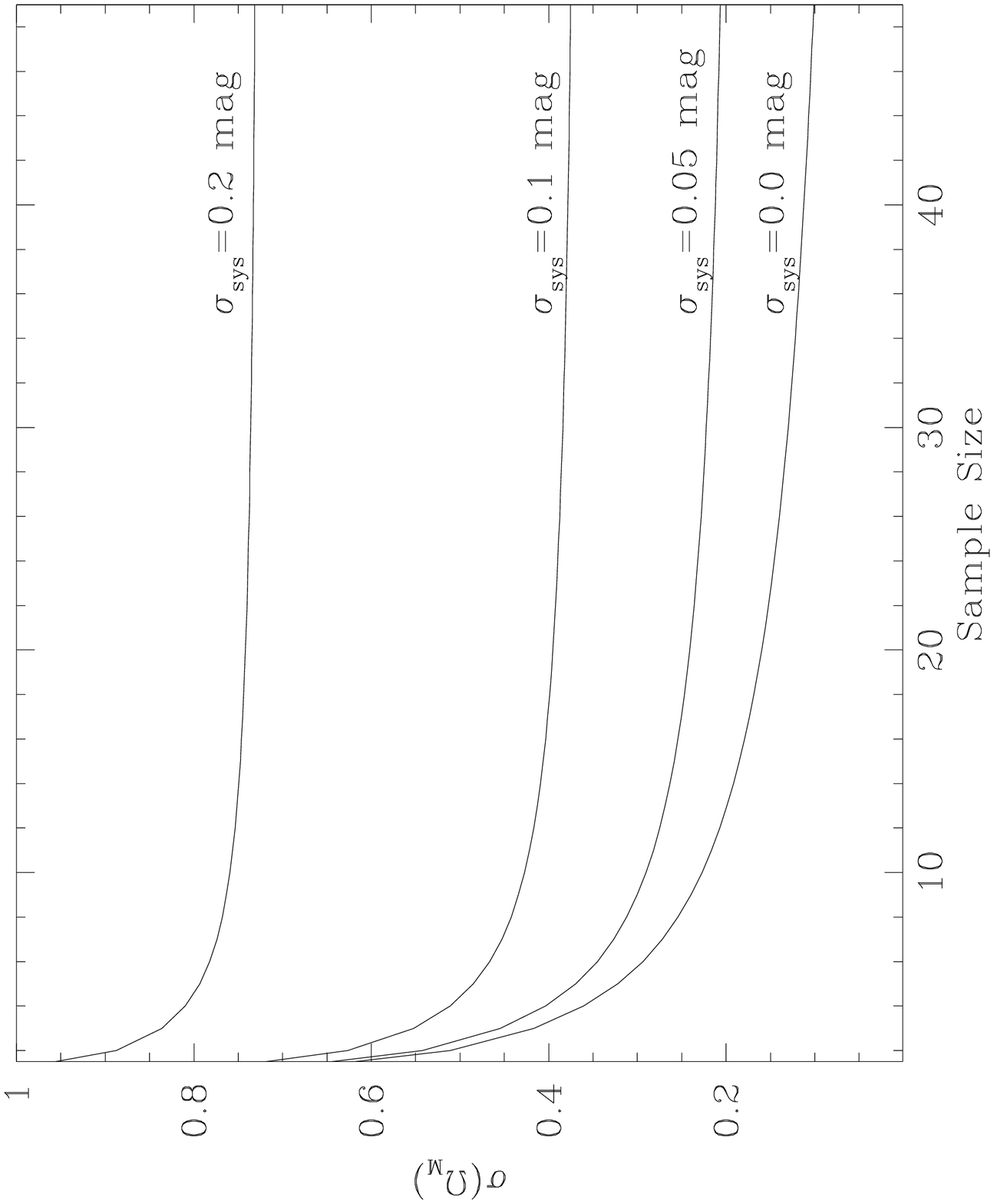]{The uncertainty in $\Omega = \Omega_M$ as a function of the sample
size for objects with observations similar to SN~1995K observed at $z=0.5$. Four
cases of systematic error are considered [$\sigma_{sys}=0.0$, 0.05 (expected), 0.10 mag,
and 0.20 mag] and are plotted as separate lines in the figure.\label{fig:future}}

\figcaption[fig13.ps]{Predicted error contours (68.3\% and 95.4\% joint confidence) in measuring
$\Omega_M$ and $\Omega_\Lambda$ simultaneously for  a sample of 30 $z=0.5$ SN~Ia augmented with
10 $z=1$ SN~Ia observed with {\it HST}.  The underlying cosmology in the simulation
is $\Omega_M=0.4,\Omega_\Lambda=0$. Also shown, as dotted lines, are the predicted
uncertainty contours (68.3\% and 95.4\% confidence) from future CMB anisotropy measurements
to be made by the Planck Surveyor mission (Zaldarriaga, Spergel, \& Seljak 1997). \label{fig:future1}}

\figcaption[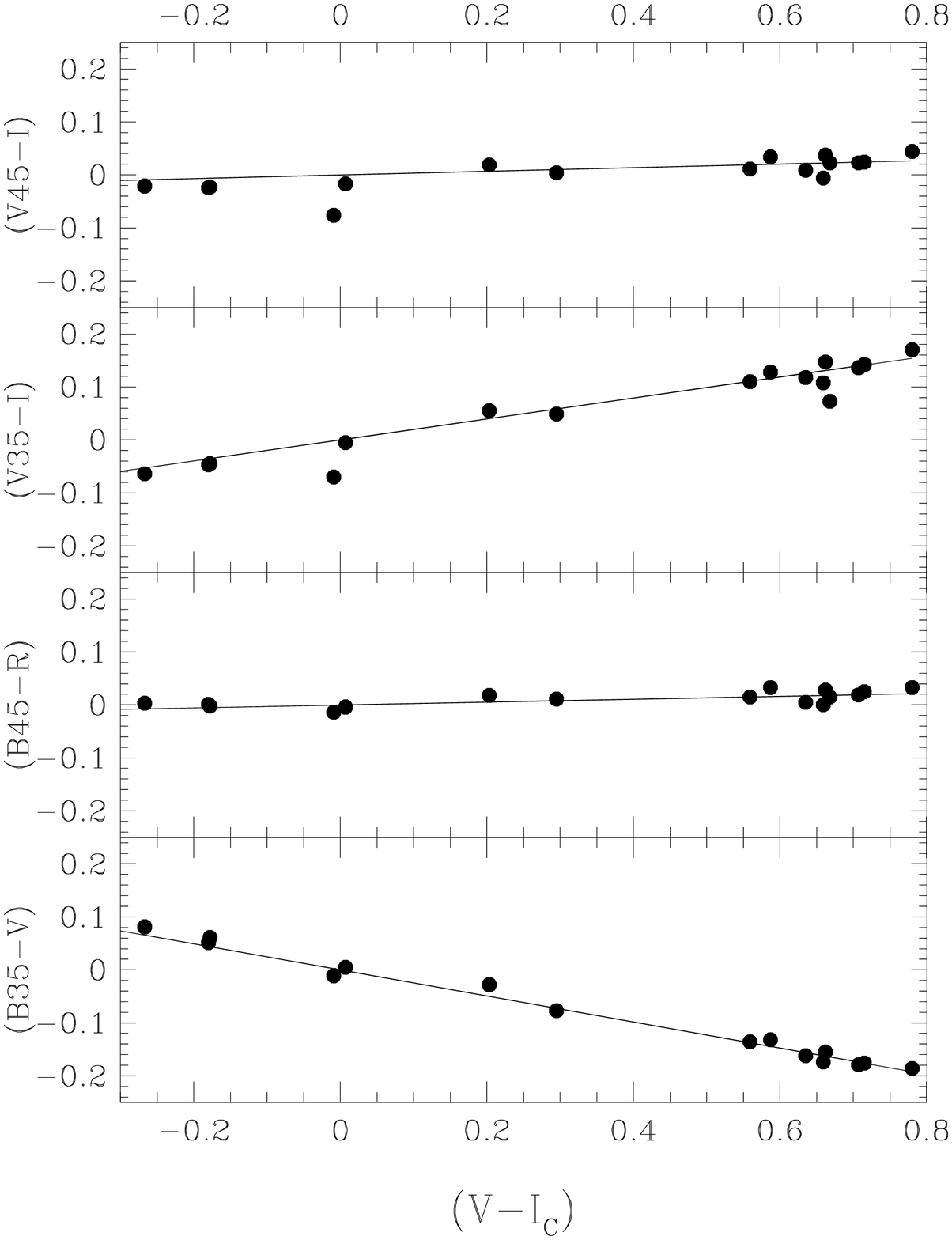]{ $(B35-V)$, $(B45-R_C)$, $(V35-I_C)$, and $(V45-I_C)$ plotted as a function of
$(V-I_C)$. The best fitting linear transformation is superposed as given in Table \ref{tab:stds}.\label{fig:HZphot}} 
\newpage
\begin{deluxetable}{rc|rc}
\small
\tablewidth{6in}
\tablecaption{Summary of Error Contributions to High Supernova Distances\label{tab:sys}}
\tablehead{\colhead{Systematic Uncertainties($1\sigma$)}& \colhead{(mag)} &\colhead{Statistical Uncertainties} &\colhead{(mag)} \\ 
}
\startdata
Photometric System Zero Point\tablenotemark{a}& 0.05   	&	Individual Zero Points  & 0.02 \nl 
Selection Effects	      & 0.02   	&	Shot noise	        & 0.15 \nl
Evolution		      & $<0.17$	& K-corrections 		& 0.03 \nl
Evolution of Extinction Law   & 0.02   	& Extinction		& 0.10 \nl
Gravitational Lensing	      & 0.02   	& $\sigma$ of SN~Ia     & 0.15 \nl 
\enddata 
\tablenotetext{a}{Includes propogated effect on extinction, $3.1 \sigma E(B-V)$.} 
\end{deluxetable}

\newpage
\begin{deluxetable}{lcccc} 
\tablecaption{A Photometric Sequence Near SN~1995K\label{tab:95Kseq}}
\tablehead{\colhead{Star}&\colhead{B45}&\colhead{V45}&\colhead{R}&\colhead{I}\\}
\startdata
1 & 21.72(03) & 20.91(04)& 21.66(03) & 20.85(04)\nl
2 & 20.06(01) & 19.60(01)& 20.03(02) & 19.57(03)\nl
3 & 17.05(01) & 16.40(01)& 17.01(02) & 16.36(02)\nl
4 & 20.51(02) & 18.88(04)& 20.40(03) & 18.77(04)\nl
5 & 17.28(02) & 16.51(03)& 17.22(03) & 16.45(04)\nl
6 & 17.18(02) & 16.83(03)& 17.15(03) & 16.80(04)\nl
7 & 20.72(03) & 20.31(03)& 20.69(04) & 20.28(05)\nl
8 & 19.75(02) & 19.40(03)& 19.73(03) & 19.37(03)\nl
9 & 18.16(02) & 17.72(03)& 18.13(03) & 17.69(03)\nl
10 &20.64(04) & 19.11(03)& 20.53(04) & 19.01(03)\nl
11 &18.57(02) & 17.60(02)& 18.50(02) & 17.54(03)\nl
12 &19.83(02) & 19.46(03)& 19.81(03) & 19.44(03)\nl
13 &19.63(02) & 17.99(03)& 19.52(03) & 17.88(03)\nl
\enddata
\tablecomments{Uncertainties in hundredths of a magnitude are listed in parentheses.}
\end{deluxetable}

\begin{deluxetable}{llccccll} 
\footnotesize
\tablecaption{Photometric Data for SN~1995K\label{tab:95Klc}}
\tablehead{\colhead{JD}&\colhead{Date}&\colhead{B45}&\colhead{V45}&\colhead{R}&\colhead{I}&\colhead{Telescope}&\colhead{Observer}}
\startdata
2449774.6&1995~Feb~26& $>$24.5  & \nodata  &\nodata  &	\nodata  & CTIO~4m  &  Hamuy et al.   \nl
2449783.6&1995~Mar~07& 24.09(38)& \nodata  &\nodata  &	\nodata  & CTIO~4m  &  Hamuy et al.   \nl
2449801.7&1995~Mar~25& 22.26(15)& \nodata  &\nodata  &	\nodata  & CTIO~4m  &  Hamuy et al.   \nl
2449806.6&1995~Mar~30& 22.19(06)& \nodata  &\nodata  &	\nodata  & CTIO~4m  &  Hamuy et al.   \nl
2449810.6&1995~Apr~03& \nodata  & 22.02(30)&22.37(12)& 	\nodata  & ESO~NTT  &  Leibundgut \& Spyromilio \nl
2449812.6&1995~Apr~05& \nodata  & \nodata  &22.28(15)& 	\nodata  & ESO~3.6m  &  Walsh       \nl
2449812.6&1995~Apr~05& 22.23(09)& 22.18(10)&\nodata  &	\nodata	 & LCO~2.4m &  Dressler    \nl
2449814.7&1995~Apr~07& 22.44(14)& 22.25(15)&\nodata  &	\nodata  & LCO~2.4m &  Dressler    \nl
2449815.6&1995~Apr~08& 22.39(12)& 22.48(22)&\nodata  &	\nodata  & LCO~2.4m &  Dressler    \nl
2449815.7&1995~Apr~08& \nodata  & \nodata  &22.64(14)&	22.36(14)& KPNO~4m  &  Ciardullo   \nl
2449816.6&1995~Apr~09& 22.62(12)& 22.48(12)&\nodata  &	\nodata  & LCO~2.4m &  Dressler    \nl
2449817.5&1995~Apr~10& 22.79(12)& \nodata  &\nodata  &	\nodata  & LCO~2.4m &  Dressler    \nl
2449830.6&1995~Apr~23& 23.62(30)& 22.95(30)&\nodata  &	\nodata  & ESO~NTT  &  Leibundgut  \nl
2449831.5&1995~Apr~24& \nodata  & \nodata  &23.70(25)\tablenotemark{a}&	22.95(30)\tablenotemark{b}& ESO~NTT  &  Leibundgut  \nl
2449836.5&1995~Apr~29& 23.96(33)& 22.92(29)&\nodata  &	\nodata  & ESO~1.5m &  Leibundgut  \nl
2449866.4&1995~May~29& $>$24.30 & $>$23.50    &\nodata  &	\nodata  & ESO~NTT  &  Spyromilio  \nl
\enddata 
\tablecomments{Uncertainties in hundredths of a magnitude are listed in parentheses.}
\tablenotetext{a}{Gunn-$r$.}
\tablenotetext{b}{Gunn-$i$.}
\end{deluxetable}

\begin{deluxetable}{lrccccll} 
\tablecaption{The Light Curve of  SN~1995K Corrected to the Restframe\label{tab:95Ktab}}
\tablehead{\colhead{JD}&\colhead{age(days)}&\colhead{B}&\colhead{V}&\colhead{$K_B$}&\colhead{$K_V$}}
\startdata
2449774.6&-20.8&$>25.23$   &\nodata    &-0.73&\nodata   \nl
2449783.6&-14.7 &24.82(38) &\nodata    &-0.73&\nodata   \nl
2449801.7& -2.4 &23.01(15) &\nodata    &-0.75&\nodata   \nl
2449806.6&  0.9 &22.94(06) &\nodata    &-0.75&\nodata   \nl
2449810.6&  3.6 &23.11(12) &22.87(30)  &-0.74&-0.85 \nl
2449812.6&  4.9 &23.02(15) &\nodata    &-0.74&\nodata   \nl
2449812.6&  4.9 &22.96(09) &23.02(10)  &-0.73&-0.84 \nl
2449814.7&  6.4 &23.18(14) &23.09(15)  &-0.74&-0.84 \nl
2449815.6&  7.0 &23.13(12) &23.32(22)  &-0.74&-0.84 \nl
2449815.7&  7.0 &23.36(14) &23.22(14)  &-0.72&-0.86 \nl
2449816.6&  7.6 &23.35(12) &23.32(12)  &-0.73&-0.84 \nl
2449817.5&  8.2 &23.53(12) &\nodata    &-0.74&\nodata   \nl
2449830.6& 17.1 &24.34(28) &23.78(27)  &-0.72&-0.83 \nl
2449831.5& 17.7 &24.43(25) &23.72(30)  &-0.73&-0.77 \nl
2449836.5& 21.1 &24.69(33) &23.74(29)  &-0.73&-0.82 \nl
2449866.4& 41.3 &$>25.02$  &$>24.31$   &-0.72&-0.81 \nl
\enddata 

\tablecomments{Uncertainties in hundredths of a magnitude are listed in parentheses.}
\end{deluxetable}

\begin{deluxetable}{lllll} 
\tablecaption{Sensitivity Functions of the High-Z Filter Set\label{tab:sens}}
\tablehead{\colhead{Wavelength(\AA)}&\colhead{$S_\lambda$(B35)}&\colhead{$S_\lambda$(B45)}&\colhead{$S_\lambda$(V35)}&\colhead{$S_\lambda$(V45)}} 
\startdata
5100.0& 0.002&	0.000&	 0.000&	 0.000 \nl
5200.0& 0.246&	0.000&	 0.000&	 0.000 \nl
5300.0& 0.708&	0.000&	 0.000&	 0.000 \nl
5400.0& 0.923&	0.000&	 0.000&	 0.000 \nl
5500.0& 0.996&   0.000&	 0.000&	 0.000\nl
5600.0& 1.000&   0.127&	 0.000&	 0.000\nl
5700.0& 0.987&   0.628&	 0.000&	 0.000\nl
5800.0& 0.953&   0.933&	 0.000&	 0.000\nl
5900.0& 0.904&   0.930&	 0.000&	 0.000\nl
6000.0& 0.938&   0.962&	 0.000&	 0.000\nl
6100.0& 0.964&   0.987&	 0.000&	 0.000\nl
6200.0& 0.884&   0.909&	 0.000&	 0.000\nl
6300.0& 0.878&   0.944&	 0.000&	 0.000\nl
6400.0& 0.924&   1.000&	 0.000&	 0.000\nl
6500.0& 0.864&   0.934&  0.000&  0.000\nl
6600.0& 0.702&   0.917&  0.128&	 0.000\nl
6700.0& 0.440&   0.927&  0.606&	 0.000\nl
6800.0& 0.237&   0.887&  0.868&	 0.000\nl
6900.0& 0.121&   0.857&  0.950&	 0.000\nl
7000.0& 0.064&   0.795&  0.965&  0.000\nl
7100.0& 0.041&   0.573&  1.000&  0.080\nl
7200.0& 0.027&   0.330&  0.962&  0.329\nl
7300.0& 0.017&   0.180&  0.892&  0.649\nl
7400.0& 0.013&   0.095&  0.861&  0.860\nl
7500.0& 0.009&   0.055&  0.837&  0.978\nl
7600.0& 0.007&   0.036&  0.817&  1.000\nl
7700.0& 0.004&   0.023&  0.832&  0.952\nl
7800.0& 0.002&   0.017&  0.849&  0.908\nl
7900.0& 0.000&   0.013&  0.814&  0.882\nl
8000.0&0.000 &  0.011 &  0.578&  0.855\nl
8100.0&0.000 &  0.000 &  0.323&  0.849\nl
8200.0&0.000 &	0.000 &	 0.168&  0.829 \nl
8300.0&0.000 &	0.000 &	 0.088&   0.719 \nl
8400.0&0.000 &	0.000 &	 0.048&   0.500 \nl
8500.0&0.000 &	0.000 &	 0.029&   0.312 \nl
8600.0&0.000 &	0.000 &	 0.018&   0.171 \nl
8700.0&0.000 &	0.000 &	 0.012&   0.097 \nl
8800.0&0.000 &	0.000 &	 0.009&   0.054 \nl
8900.0&0.000 &	0.000 &	 0.006&   0.034 \nl
9000.0&0.000 &	0.000 &	 0.005&   0.021 \nl
9100.0&0.000 &	0.000 &	 0.003&   0.014 \nl
9200.0&0.000 &	0.000 &	 0.002&   0.009 \nl
9300.0&0.000 &	0.000 &	 0.001&   0.006 \nl
9400.0&0.000 &	0.000 &	 0.000&   0.003 \nl
9500.0&0.000 &	0.000 &	 0.000&   0.001 \nl
\enddata 
\end{deluxetable}

\begin{deluxetable}{lrrrr} 
\tablecaption{Primary Standards for the High-Z Standard System \label{tab:stds}}
\tablehead{\colhead{Star Name\tablenotemark{a}}&\colhead{B35}&\colhead{B45}&\colhead{V35}&\colhead{V45}} 
\startdata
    HR718 & 4.285&   4.299 &  4.329 &  4.340  \nl
    HR1544& 4.352&   4.344 &  4.340 &  4.333  \nl
    HR3454& 4.346&   4.378 &  4.448 &  4.480  \nl
    HR4468& 4.711&   4.721 &  4.749 &  4.760  \nl
    HR4963& 4.373&   4.369 &  4.373 &  4.370  \nl
    HR5501& 5.699&   5.696 &  5.699 &  5.699  \nl
    HR7596& 5.583&   5.548 &  5.499 &  5.473  \nl
    HR7950& 3.765&   3.761 &  3.764 &  3.764  \nl
    HR8634& 3.428&   3.439 &  3.474 &  3.489  \nl
    HR9087& 5.147&   5.163 &  5.207 &  5.228  \nl
      L377& 11.077&  10.947&  10.750&  10.656\nl
     L1020& 11.335&  11.180&  10.938&  10.820 \nl
      EG21& 11.431&  11.468&  11.513&  11.536 \nl
     L1788& 12.991&  12.854&  12.636&  12.527 \nl
     L2415& 12.074&  11.955&  11.761&  11.662 \nl
      H600& 10.380&  10.326&  10.245&  10.200 \nl
     L3218& 11.818&  11.776&  11.698&  11.662 \nl
     L3864& 12.009&  11.870&  11.649&  11.539 \nl
     L4364& 11.435&  11.362&  11.266&  11.221 \nl
       F56& 11.072&  11.085&  11.126&  11.141 \nl
     L4816& 13.787&  13.791&  13.770&  13.758 \nl
      CD32& 10.365&  10.290&  10.185&  10.135 \nl
     L6248& 11.638&  11.493&  11.261&  11.147 \nl
     EG274& 11.093&  11.140&  11.215&  11.258 \nl
     L7379& 10.048&   9.892&   9.656&   9.543 \nl
     L7987& 12.278&  12.311&  12.350&  12.372 \nl
     L9239& 11.876&  11.711&  11.452&  11.326 \nl
      F110& 11.893&  11.957&  12.081&  12.138 \nl
     L9491& 14.093&  14.067&  14.043&  14.037 \nl
\enddata
\tablenotetext{a}{As listed in Hamuy et al. (1992).}
\end{deluxetable}

\begin{deluxetable}{ll} 
\tablecaption{The High-Z Filter System Transformations and Zero Points\label{tab:stdsi}}
\tablehead{\colhead{Transformation}&\colhead{Zero Point\tablenotemark{a}}} 
\startdata
$B35 = -0.246(V-I_C)+{\rm V}$  &	ZP$=-21.339$ \nl
$B45 = +0.027(V-I_C)+{\rm R_C}$&	ZP$=-21.582$ \nl
$V35 =+0.198(V-I_C)+{\rm I_C}$ &	ZP$=-22.045$ \nl
$V45 = +0.034(V-I_C)+{\rm I_C}$&	ZP$=-22.292$ \nl
\enddata
\tablenotetext{a}{Zero Point required for equation (\ref{eq:magz=0}) if $F_\lambda$ is in units of ergs~cm$^{-2}$~s$^{-1}$~\AA$^{-1}$.}
\end{deluxetable}


\begin{references}
\reference{A69} Arnett, W. D. 1969, Ap\&SS, 5, 280
\reference{B89} Barbon, R., Rosino, L., \& Iijima, T. 1989, A\&A, 220, 83
\reference{B57} Baum, W.A. 1957,  AJ, 62, 6 
\reference{B90} Bessell, M.S. 1990, PASP, 102, 1181
\reference{Bouchet85} Bouchet, P., Lequeux, J., Maurice, E., Prevot, L., Prevot-Burnichon, M.L., 1985, A\&A, 149, 330.
\reference{B87} Branch, D. 1987, ApJ, 316, 81L
\reference{B96} Branch, D., Fisher, A., Baron, E., \& Nugent, P. 1996, ApJ, 470, L7 
\reference{BM93} Branch, D., \& Miller, D.L. 1993,  ApJ, 405, L5
\reference{C97} Cappellaro, E. et al. 1997, A\&A, 322, 431
\reference{CL95} Coles, P., \& Lucchin, F. 1995, ``Cosmology'' (John Wiley \& Sons: Chicester), pp. 31-46
\reference{C80} Cousins, A.W.J. 1980, MNSSA, 39, 93
\reference{CM69} Colgate, S., \& McKee, W., 1969, ApJ, 157, 623
\reference{E94} Evans, R. 1994, PASAu, 11, 7
\reference{F97} Filippenko, A. V. 1997, ARAA, 35, 309 
\reference{F92a} Filippenko, A. V., et al. 1992a, ApJ, 384, L15
\reference{F92b} Filippenko, A. V., et al. 1992b, AJ, 104, 1543
\reference{F95} Fisher, A., Branch, D., H\"oflich, P., \& Khokhlov, A. 1995, ApJ, 447, L73
\reference{Ford93} Ford, C. H., Herbst, W., Richmond, M.W., Baker, M.L., Filippenko, A.V., Treffers, R.R., Paik, Y., \& Benson, P.J. 1993, AJ, 106, 1101
\reference{G96a} Garnavich, P.,  et al. 1996a, IAUC 6332
\reference{G96b} Garnavich, P.,  et al. 1996b, IAUC 6358
\reference{G97a} Garnavich, P.,  et al. 1997a, IAUC 6633
\reference{G98} Garnavich, P. et al. 1998b, ApJ, 000, L000
\reference{GS97} Gonzalez, A. H., \& Faber, S. M. 1997, ApJ, 485, 80
\reference{G95} Goobar, A., \& Perlmutter S. 1995, ApJ, 450, 14
\reference{G86} Groth, E.J. 1986, AJ, 91, 1244
\reference{DGL98} Guerra, E.J., \& Daly, R. A. 1998, ApJ, 493, 536 
\reference{G78} Gunn, J.E. 1978 in Observational Cosmology: The 8th Advanced Course
of the Swiss Society of Astronomy and Astrophysics, ed. A Maeder, L. Martinet,
\& G. Tammann (Saverny: Geneva), p 1.
\reference{H93a} Hamuy, M., et al. 1993a, AJ, 106, 2392
\reference{H93b} Hamuy, M., Phillips, M. M., Wells, L. A., \& Maza, J. 1993b, PASP, 105, 787
\reference{H94} Hamuy et al. 1994, AJ, 108, 2226
\reference{H95} Hamuy, M., Phillips, M.M., Maza, J., Suntzeff, N.B., Schommer, R.A., \& Avil\'es, R. 1995, AJ, 109, 1 [H95]
\reference{H96a} Hamuy, M., Phillips, M.M., Schommer, R.A., Suntzeff, N.B., Maza, J., \& Avil\'es, R. 1996, AJ, 112, 2391 [H96a]
\reference{H96b} Hamuy, M., Phillips, M.M., Suntzeff, N.B., Schommer, R.A., Maza, J., \& Avil\'es, R. 1996, AJ, 112, 2398 [H96b]
\reference{H96c} Hamuy, M., et al. 1996, AJ, 112, 2408 [H96c]
\reference{H96d} Hamuy, M., Phillips, M.M., Suntzeff, N.B., Schommer, R.A., Maza, J., Smith, R.C., Lira, P., \& Avil\'es, R. 1996, AJ, 112, 2438 [H96d]
\reference{HKW95} H\"oflich, P., Khokhlov, A., \& Wheeler, J.C. 1995, ApJ, 444, 831
\reference{HWT98} H\"oflich, P., Wheeler, J.C., \& Thielemann, F. K. 1998, ApJ, 495, 617
\reference{HW97} Holz, D. E., Wald, R. M., 1998, astro-ph/9708036
\reference{HF60} Hoyle, F., \& Fowler, W.A. 1960, ApJ, 132, 565
\reference{Hu96} Hu, W. 1996, in The Universe at High z, Large-Scale Structure, and the Cosmic Microwave Background, ed. E. Martínez-González \& J. L. Sanz (Berlin: Springer), 207
\reference{HMS56} Humason, M.L., Mayall, N.U., \& Sandage, A.R. 1956,  ApJ,  61, 97
\reference{KVB95} Kantowski, R., Vaughan, T., \& Branch, D. 1995, ApJ, 447, 35
\reference{K93} Kellerman, K.I., 1993, Nature, 361, 134
\reference{K92} Kennicutt, R.C., 1992, ApJS, 79, 255 
\reference{K93} Khokhlov, A., M\"uller, E., \& H\"oflich, P. 1993, A\&A, 270, 223
\reference{K95} Kirshner, R. P., et al. 1995, IAUC No. 6267
\reference{K96} Kim, A., Goobar, A., \& Perlmutter, S. 1996, PASP, 108, 190
\reference{K97} Kim, A., et al. 1997, ApJ, 483, 565
\reference{KB97} Kolatt, T.S., \& Bartelmann, M. 1998 MNRAS, submitted Astro-ph/9708120
\reference{K68} Kowal, C. T. 1968,  AJ,  73, 1021
\reference{L92} Landolt, A. U. 1992a, AJ, 104, 340
\reference{L92a} Landolt, A. U. 1992b, AJ, 104, 372
\reference{LP92} Lauer, T., \& Postman, M. 1992, ApJ, 400, 47
\reference{L90} Leibundgut, B. 1990, A\&A, 229, 1
\reference{LS97} Leibundgut, B., \& Spyromilio, J. 1997 in The Early Universe with the VLT, 
ed. J. Bergeron, (Berlin: Springer), 95
\reference{L93} Leibundgut, B., et al. 1993, AJ, 105, 301
\reference{L96} Leibundgut, B., et al. 1996, ApJ, 466, L21
\reference{L96} Lira, P. 1996, M.S. thesis, Universidad de Chile

\reference{L90} Livne, E. 1990, ApJ, 354, L53
\reference{M90} McNaught, R. H. 1990, IAUC 5039
\reference{MWW97} Martin, R., Williams, A., \& Woodings, S. 1997, IAUC 6558
\reference{M58} Mattig, W. 1958, Astr.Nach., 184, 109 
\reference{M60} Minkowski, R. 1960, ApJ, 132, 908
\reference{M89} Mueller, J. 1989, IAUC 4920
\reference{NTY84} Nomoto, K., Thielemann, F., \& Yokoi, K. 1984, ApJ, 286, 644
\reference{NN89} N\o rgaard-Nielsen, H. U., et al. 1989, Nature, 339, 523
\reference{O79} Oemler, A., \& Tinsley, B. M. 1979,  AJ, 84, 985
\reference{OS68} Oke, J.B., \& Sandage, A. 1968, ApJ, 154, 21
\reference{OGH96} Oke, J.B., Hoessel, J. E., \& Gunn, J. G. 1996, AJ, 111, 29
\reference{Peebles93} Peebles, P. J. E. 1993, ``Principles Of Physical Cosmology'', Princeton University Press, Princeton, New Jersey 
\reference{Pe92} Perlmutter, S., et al. 1992 in ``Robotic Telescopes in the 1990s'' ed. A.V. Filippenko (San Francisco: ASP Conf. Ser. 34), p. 67
\reference{Pe95} Perlmutter, S., et al. 1995, ApJ, 440, L41 
\reference{Pe97} Perlmutter, S., et al. 1997, ApJ, 483, 565
\reference{Pe98} Perlmutter, S., et al. 1998, Nature, 391, 51 (Erratum: 392, 311)
\reference{PD95} Phillips, A.C., \& Davis, L.E. 1995 ``Astronomical Data Analysis Software and Systems'',
 ed. R.A. Shaw, H.E. Payne, \& J.J.E. Hayes (San Francisco:  ASP Conf. Ser. 77), p. 297
\reference{P92} Phillips, M. M., Wells, L.A., Suntzeff, N.B., Hamuy, M., Leibundgut, B., Kirshner, R. P., \& Foltz, C.B. 1992, AJ, 103, 1632
\reference{P93}Phillips, M. M. 1993, ApJ, 413, L105 [P93]
\reference{P98}Phillips, M. M., et al. 1998, in preparation
\reference{P87}Phillips, M. M., et al. 1987, PASP, 99, 592
\reference{P92} Pollas, C. 1992, IAUC 5420
\reference{Reiss97} Reiss, D., Germany, L., Schmidt, B. P., \& Stubbs, C. 1998, AJ, 115, 26.
\reference{RPK95} Riess, A.G., Press, W. H., \& Kirshner, R. P.  1995, ApJ, 438, L17 [RPK95]
\reference{RPK96} Riess, A.G., Press, W. H., \& Kirshner, R. P.  1996a, ApJ, 473, 88 [RPK96]
\reference{RPK96b} Riess, A.G., Press, W. H., \& Kirshner, R. P.  1996b, ApJ, 473, 588
\reference{retal97} Riess, A.G., et al. 1997, AJ, 114, 722
\reference{R98} Riess, A.G., et al. 1998, AJ submitted
\reference{R36} Robertson, H.P. 1936,  ApJ, 83, 187
\reference{RP92} Ruiz-Lapuente, P., Cappellaro, E., Turatto, M., Gouiffes, C., Danziger, I.J., Della Valle, M., \& Lucy, L.B. 1992, ApJ, 387, L33
\reference{Rh97} Rhoads, J. E. 1997, submitted to ApJ astro-ph/9705163
\reference{Setal96}Saha, A., Sandage, A., Labhardt, L., Tammann, G. A.,Macchetto, F. D., \& Panagia, N. 1997, ApJ, 486, 1
\reference{S61}	 Sandage, A.R., 1961, ApJ, 133, 355
\reference{ST93} Sandage, A., \& Tammann, G.A. 1993, ApJ, 415, 1
\reference{SMS93} Schechter, P.L., Mateo, M., \& Saha, A. 1993, PASP, 105, 1342
\reference{S97a} Schmidt, B. P. 1997, in ``Thermonuclear Supernovae'', ed. P. Ruiz-Lapuente, R. Canal, \& J. Isern (Dordrecht: Kluwer), p. 765
\reference{S95} Schmidt, B., et al. 1995, IAUC 6160
\reference{S96} Schmidt, B., et al. 1996, BAAS, 28, 1420
\reference{S97a} Schmidt, B., et al. 1997a, IAUC 6602
\reference{S97b} Schmidt, B., et al. 1997b, IAUC 6646
\reference{S84} Shanks, T., et al. 1984, MNRAS,  206, 767
\reference{SS95} Stepanas, P.G., \& Saha, P. 1995, MNRAS, 272, L13
\reference{Stet87} Stetson, P. B. 1987, PASP, 99, 191
\reference{Sun96} Suntzeff, N.B., et al. 1996, IAUC 6490
\reference{TS95} Tammann, G.A., \& Leibundgut, B. 1990,  A\&A, 236, 9
\reference{TL90} Tammann, G.A., \& Sandage, A. 1995,  ApJ, 452, 16
\reference{T72} Tinsley, B. 1972, ApJ, 178, 319
\reference{T93} Treffers, R., Leibundgut, B., Filippenko, A.V., \& Richmond, M. W. 1993, BAAS, 25, 834
\reference{Tu96} Turatto, M., et al. 1996, MNRAS, 283, 1
\reference{v95} van den Bergh, S. 1995, ApJ, 453, L55
\reference{vBP92} van den Bergh, S., \& Pazder, J. 1992, ApJ, 390, 34
\reference{v97} von Hippel, T., Bothun, G. D., \& Schommer, R.A. 1997, AJ, 114, 1154  
\reference{W36} Walker, A.G. 1936, Proc. Lond. Math. Soc., 42, 90
\reference{W97} Wambsganss, J., Cen, R., Guohong, X., \& Ostriker, J. 1997, ApJ, 475, L81
\reference{W72} Weinberg, S. 1972, Gravitation and Cosmology: Principles and Applications of the General Theory of Relativity (John Wiley \& Sons: New York) 
\reference{WH90} Wheeler, J.C., \& Harkness R. P. 1990, Rep. Prog. Phys. 53, 1467
\reference{WW94} Woosley, S. E., \& Weaver, T.A. 1994, ApJ, 423, 371
\reference{yp95} Yoshi, Y., \& Peterson, B. A. 1995, \apj, 444, 15
\reference{ZSS97} Zaldarriaga, M., Spergel, D.N., \& Seljak, U. 1997, ApJ, 488, 1
\reference{Z68} Zwicky, F. 1968, \pasp, 80, 462
\end{references}
\end{document}